\documentclass[manuscript]{aastex}
\usepackage{epsfig}
\usepackage{spr-astr-addons}
\usepackage{url}
\usepackage{url}
\usepackage{natbib}
\usepackage[colorlinks=true,citecolor=blue]{hyperref}
\usepackage{graphicx,color}
\usepackage{times}
\usepackage[normalem]{ulem}
\usepackage{amssymb}
\newcommand{\Msun}{\ensuremath{M_{\odot}}}
\newcommand{\Zsun}{\ensuremath{Z_{\odot}}}
\newcommand{\Ha}{H$\alpha$}

\begin{document}
\title{X-ray cavities and temperature jumps in the environment of the strong cool core cluster Abell 2390}
\shorttitle{X-Ray Cavities Abell 2390}
\shortauthors{Sonkamble et al.}
\author{S. S. Sonkamble$^{1}$, N. D. Vagshette$^{1,2}$, P. K. Pawar$^{1}$,
M.K.Patil$^{1}$\altaffilmark{*}}
\affil{$^{1}$School of Physical Sciences, Swami Ramanand Teerth Marathwada
University, Nanded 431 606, India.}
\affil{$^{2}$ Inter University Centre for Astronomy \& Astrophysics (IUCAA),
Pune 411 007, India.\\ }
\email{patil@iucaa.ernet.in}

\begin{abstract}
We present results based on the systematic analysis of high resolution 95\,ks \textit{Chandra} observations of the strong 
cool core cluster Abell 2390 at the redshift of $z$ = 0.228 that hosts an energetic radio AGN. This analysis has enabled 
us to investigate five X-ray deficient cavities in the atmosphere of Abell 2390 within central 30\arcsec. Presence of these 
cavities have been confirmed through a variety of image processing techniques like, the surface brightness profiles, unsharp 
masked image, as well as 2D elliptical model subtracted residual map. Temperature profile as well as 2D temperature map 
revealed structures in the distribution of ICM, in the sense that ICM in the NW direction is cooler than that on the SE direction. 
Temperature jump in all directions is evident near 25\arcsec\,(90.5 kpc) corresponding to the average Mach number 1.44$\pm$0.05, while 
another jump from 7.47 keV to 9.10 keV at 68\arcsec\,(246 kpc) in the north-west direction, corresponding to Mach 
number 1.22$\pm$0.06 and these jumps are associated with the cold fronts. Tricolour map as well as hardness ratio map detects 
cool gas clumps in the central 30 kpc region  of temperature $4.45_{-0.10}^{+0.16}$\,keV. The entropy profile derived from 
the X-ray analysis is found to fall systematically inward in a power-law fashion and exhibits a floor near 
12.20$\pm$2.54\,keV\, cm$^2$ in the central region. This flattening of the entropy profile in the core region confirms 
the intermittent heating at the centre by AGN. The diffuse radio emission map at 1.4\,GHz using VLA L-band data exhibits 
highly asymmetric morphology with an edge in the north-west direction coinciding with the X-ray edge seen in the unsharp mask image.
The mechanical power injected by the AGN in the form of X-ray cavities is found to be 5.94$\times$10$^{45}$ erg\,s$^{-1}$ 
and is roughly an order of magnitude higher than the energy lost by the ICM in the form of X-ray emission, confirming 
that AGN feedback is capable enough to quench the cooling flow in this cluster.

\end{abstract}



\section{Introduction}
\hspace{0.5cm} 
Superb resolution achieved by the {\it Chandra} X-ray Observatory has enabled us to investigate the thermodynamical and chemodynamical properties of the X-ray emitting intracluster medium (ICM) at greater details, including the detection of fluctuations in the surface brightness distribution of the ICM. As a result, majority of the studies in the last two decades were focused  on  the exploration  and understanding of the energetic feedback from super-massive black hole (SMBH) residing at the core of the largest gravitationally bound systems like groups and clusters of galaxies \citep[see review by][]{Fab12}. AGNs are believed to be powered by the binding energy released by the accretion of matter onto the SMBH. Now it is widely accepted that the energetic feedback from the AGN plays a significant role in the evolution of galaxies as well as the ICM, however, the form in which this energy is released is not yet fully understood.

The commonly observed signatures of the AGN interactions with the ICM are the bubbles or cavities in the X-ray surface brightness distribution \citep{dong10, Randall11, David09, Rafferty06, Pan12, Pan13}. In addition, sharp discontinuities or edges in the surface brightness associated with the shocks and cold fronts have also been observed in few of the clusters (e.g., Abell 1795, RXJ1720.1+2638, Abell 2142, Abell 754, Abell 3667, Abell 2204). These bubbles are often found in association with the cooling flow or the ``cool core'' clusters and are filled with radio emission originating from the central AGN. Some of the spectacular examples of the cool core clusters with radio bubbles are the Perseus cluster \citep{Fab03,Fab06}, M87 \citep{Form05}, Hydra A \citep{Nul05}, MS0735.6+7421 \citealp{McN05}, Abell 3581 \citealp{Can13}, HCG 62 \citealp{Gitti10}, and Abell 2052 \citep{blan01,blan09,blan11}.

Combined observations of the ICM in the radio and X-ray bands have provided us with several incidences of common appearance of cavities, shocks and ripples, confirming the widespread impact of the AGN feedback from the central SMBH \citep[e.g.][]{Bir04, Bir08, blan01, David09, Giac11}. In the AGN feedback scenario, the radio source associated with the central cD galaxy drives strong jet outflows which interact with the hot ICM and inflate lobes of radio-synchrotron emission. As a consequence, the gas gets displaced and results in the formation of buoyantly rising bubbles. Cavities or bubbles are nothing but the depressions in the X-ray surface brightness consisting of low density relativistic plasma, that float upward in the cooling flow atmosphere until they reach equilibrium at some large radius where the ambient entropy is equal to that within the bubble \citep{Rafferty06}.

The jet-blown cavities in the X-ray atmosphere of the clusters can act as calorimeter and hence provide a direct probe to estimate the energy input by the AGN jet in the form of mechanical feedback. Systematic studies employing deep observations with high resolution {\it Chandra} X-ray telescope have demonstrated that AGN can inject upto about 10$^{58}$-10$^{62}$ erg per outburst into the ICM \citep[see][for review]{Mc07}. This amount of mechanical  energy injected by the AGN feedback is sufficient to heat as well as quench the cooling  of the ICM on the cluster scales \citep{Bir04, Rafferty06}, and whose balance can be checked by comparing it with the energy lost by the ICM in the form of X-ray emission. To better understand the interactions between the AGN feedback and the surrounding ICM; and to confirm the balance between mechanical power fed by the AGN versus that lost by the ICM through the radiative loss, it is important to explore X-ray bright systems with apparent signatures of such interactions. Abell 2390, as it hosts a powerful radio source at its core, is a potential candidate to investigate such a balance between the AGN feedback and radiative loss by the ICM.

In this paper we present 95\,ks \textit{Chandra} observations of the cool core cluster Abell 2390 (ObsID 4193). This is a moderately rich cluster hosting about 216 members at a redshift of $z$ = 0.228 \citep{Abr96, Yee96}. Its central dominant galaxy, PGC 140982, hosts a strong and complex radio source (B2151+174, \citealp{Augusto06} and \citealp{Egami06}). Deep optical broadband images of the central dominant galaxy PGC 140982 exhibit a wealth of structures including a significant amount of dust and molecular gas surrounding the cD galaxy (\citealp{Pipino09}, \citealp{Bardeau07}) and appears very luminous in the \Ha\,\,and IR bands \citep{Raw12}. The star formation rate estimated in the core of this cluster using the \Ha\,\,flux is found to be 15 $\Msun$/yr \citep{Haines12}. ICM in Abell 2390 exhibit a strong peak in the X-ray surface brightness profile \citep{Allen01} and also exhibit several other substructures such as X-ray cavities and breaks in the X-ray surface brightness profile (\citealp{Allen01}, \citealp{Vikhlinin05}). The central cooling time of the ICM in this cluster is $t_{cool}$=1.9 Gyr with $r_{cool}$ = 60.91 kpc \citep{hla11}.

This paper is organized as follows: Section~\ref{obs} describes the data sets used for the present study and its reduction procedure, Sections~\ref{img} and \ref{spec} discuss the imaging and spectral analysis of the data set, Section~\ref{disc} discussion of our results, while Section~\ref{conc} summarizes results from this study. Throughout this paper we assume the $\Lambda$CDM cosmology with $H_0$ = 71 km s$^{-1}$ Mpc$^{-1}$, $\Omega_M$ = 0.27, and $\Omega_{\Lambda}$ = 0.73. At the redshift of Abell 2390 ($z$ = 0.228) this cosmology corresponds to a scale of 1\arcsec = 3.62 kpc and the luminosity distance of the cluster $\sim$ 1125 Mpc (Table~\ref{basicpro}). All the errors reported in this paper are at 68$\%$ confidence level (i.e. 1$\sigma$) unless otherwise stated.

\section{Observations and data preparation}
\label{obs}
Abell 2390 was observed with the \textit{Chandra} Advanced CCD Imaging Spectrometer (ACIS-S) three times during November 1999 and September 2003 with effective exposure times of 10\,ks (ObsID 500), 10\,ks (ObsID 501) and 95\,ks (ObsId 4193). ObsIDs 500 and 501 were performed in the FAINT mode and were badly affected by the background flares. Therefore, we acquired level 1 event file corresponding to the ObsID 4193 and reprocessed it using the latest version of the  CIAO software (CIAO v 4.2) and calibration files CALDB 4.5.5.1 provided by the \textit{Chandra} X-ray Center. Charge transfer inefficiency (CTI) and time-dependent gain corrections were applied wherever applicable. Periods of high background flares were identified and filtered out using the 3$\sigma$ clipping of the full-chip light curve using the task \textit{lc$\_$clean} within CIAO. This resulted in the net exposure time of 88.63\,ks. 

As diffuse X-ray emission from Abell 2390 fills the image field of view, therefore standard CALDB blank sky background frames were preferred for the background correction. The appropriate blank-sky background frames were processed in a similar way as that of the event file, reprojected to the corresponding sky position and were normalized to match the count rate in the 9 - 12\,keV energy band in the observed data set. The background subtracted science frame was then divided by the exposure map generated using the \textit{mkexpmap} so as to account for the exposure variation of the data. Point sources present in the CCD field were detected using the wavelet detection tool \textit{wavdetect} within CIAO with a  detection threshold of 10$^{-6}$ and were inspected visually before excluding from the subsequent analysis. A cleaned, background subtracted 0.3 - 7 keV\textit{Chandra} image of Abell 2390 is shown in Figure~\ref{fig:raw} (\textit{left panel}). This figure has been smoothed with a 3$\sigma$ wide Gaussian. 

\begin{table}
\caption{Global Parameters of Abell 2390}
\begin{tabular}{@{}llrrrrrrr@{}}
\hline
\hline
RA$ \&$ DEC (J2000) &21:53:36.8; +17:41:44.00 \\
Mag ($B_T$) & 17.6\\
Distance (Mpc)& 1125\\ 
Ang. Distance (Mpc)& 746\\
Redshift ($z$) & 0.228\\
Radial Velocity (km s$^{-1}$) & 68353 \\ 
Morphology & Medium Compact (Zw) \\
Cluster Diameter & 20\arcmin \\
\hline
\hline
\end{tabular}
\label{basicpro}
\end{table}
\section{X-ray Imaging Analysis}
\label{img}

\subsection{Surface Brightness Profiles}

\begin{figure*}
{
\includegraphics[width=90mm,height=90mm]{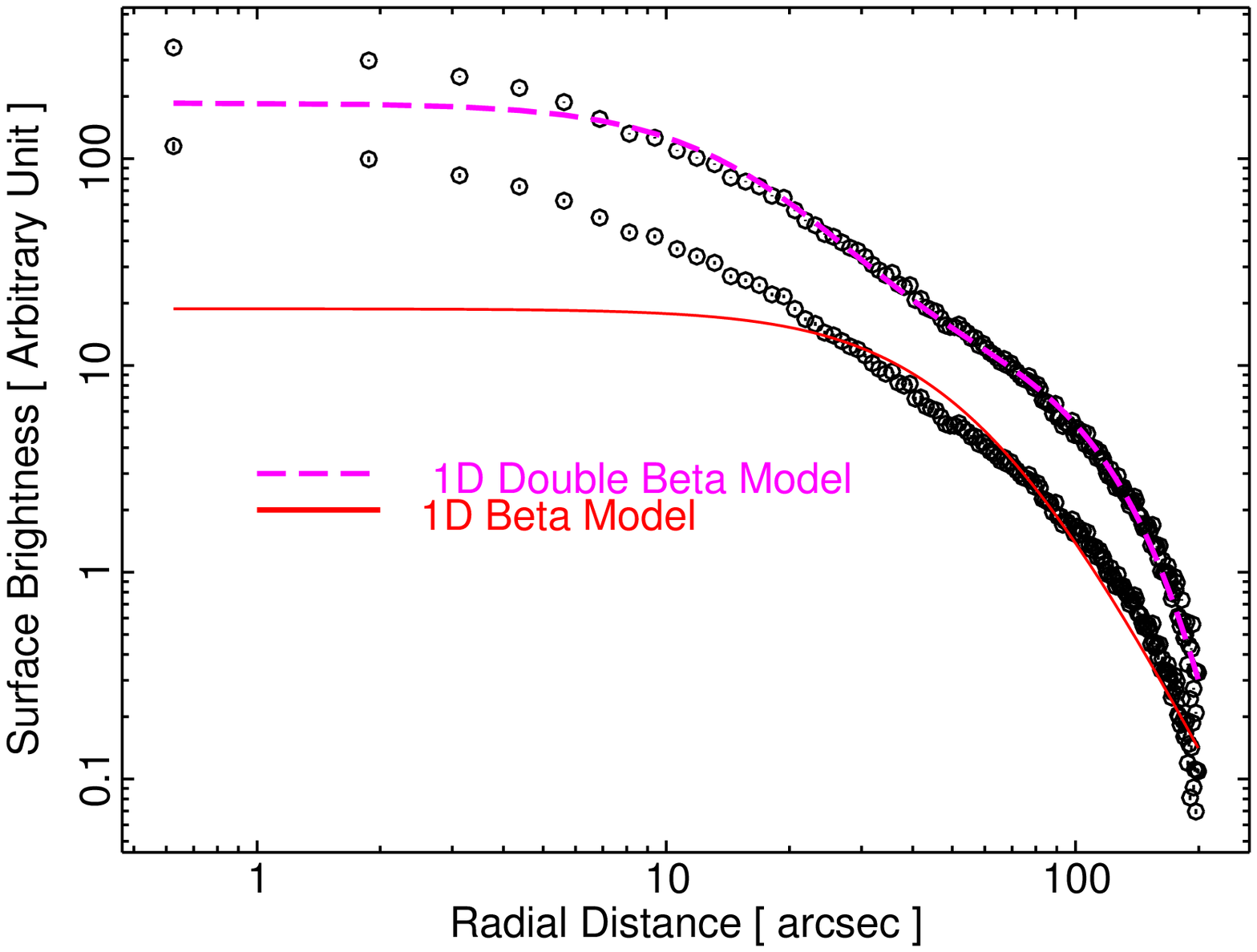}
\includegraphics[width=85mm,height=85mm]{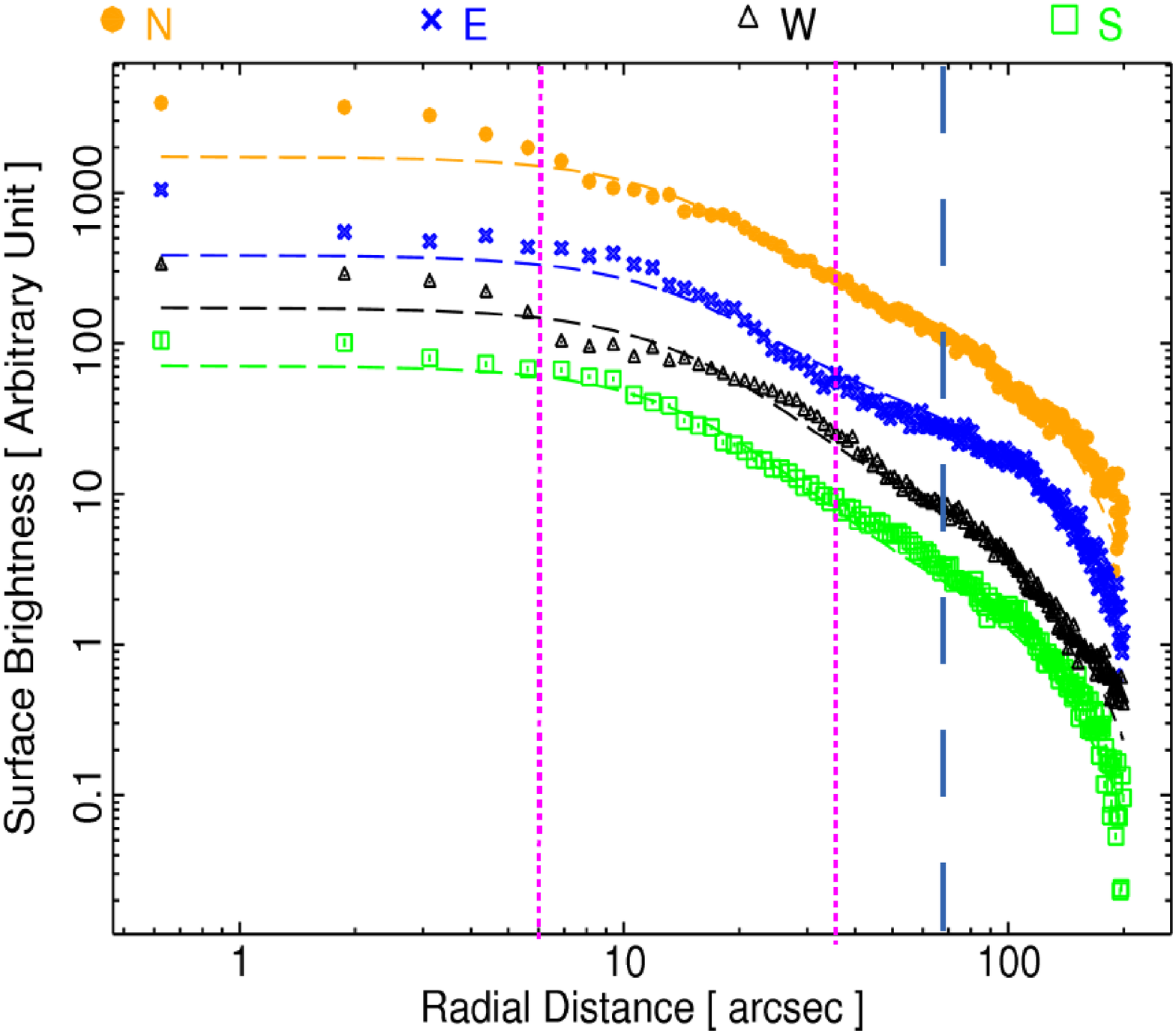}
}
\caption{{(\it left panel)}: Elliptical 0.5$-$3\,keV surface brightness profile of the X-ray emission from Abell 2390. 
The open circles represent the data points, while the continuous red line represent the fit obtained from a single $\beta$-model. 
The observed profile is not constrained properly by the single $\beta$-model, particularly in the central region. 
Therefore, a double $\beta$-model (dashed magenta line) was fitted to the data points. ( {\it right panel}): Surface brightness 
profiles of the X-ray emission extracted from 4 different sectors as marked in Figure~\ref{fig:unsharp}; 
N-sector (covers 20$^{\circ}$ -  130$^{\circ}$), E-sector (130$^{\circ}$ - 190$^{\circ}$), S-sector (190$^{\circ}$ - 290$^{\circ}$), 
and W-sector (290$^{\circ}$ - 20$^{\circ}$). For better presentation, these profiles are offset vertically by arbitrary values. 
Note the significant deviations in the regions corresponding to the X-ray cavities as well as excess emission.}
\label{fig:azmuth_beta}
\end{figure*}

\begin{table*}
\caption{Double beta model fit}
\begin{tabular}{@{}llllllccccccr@{}}
\hline\hline
{ $r_{c1}$ (arcsec)}		&{$\beta_{1}$}		& {$\sum_{01}$ (Arbitrary Unit)} 	
&{$r_{c2}$ (arcsec)}		&{$\beta_{2}$}		& {$\sum_{02}$ (Arbitrary Unit)}   \\
\hline	
524.84			&10			&4.42		
&19.04			&0.74			&57.86	\\
\hline
\end{tabular}
\footnotesize
\label{beta2d_t}
\end{table*}

In order to investigate 2D distribution of the ICM in Abell 2390, we computed surface brightness profile of the extended emission from Abell 2390 by extracting 0.5-3.0\,keV X-ray photons from a series of concentric elliptical annuli centred on the X-ray peak of the cluster emission using the task \textit{dmextract} within CIAO. The centre, ellipticity and position angle of Abell 2390 were determined by fitting a two-dimensional Lorentzian surface \citep{wise04}. The centroid of Abell 2390 was found at RA=21:53:36.82 and DEC=+17:41:43.00, while average ellipticity and position angle of the ellipses were found to be 0.36$^{+0.0042}_{-0.0042}$ and 38$\degr$, respectively. The resulting background subtracted surface brightness profile is shown in Figure~\ref{fig:azmuth_beta} (\textit{left panel}). Assuming that the X-ray emitting hot gas and galaxies in the cluster are in hydrostatic equilibrium and galaxies exhibit isothermal distribution described by the King model (\citealp[see review by][]{Gitt12}), we fit this surface brightness profile with a single $\beta$-model given by

\[ \hspace{20mm} \Sigma(r)=\Sigma_0\left[ 1+\left( \frac{r}{r_c} \right)^2\right] ^{-3\beta+0.5}\]

where $\Sigma_0$, $\Sigma(r)$, $r_c$ and $\beta$ represent the central X-ray surface brightness, surface brightness at projected distance $r$, core radius and slope parameter, respectively. The fitting was performed using the {\sc Sherpa} model \textit{beta1d}\, in conjunction with the $\chi^2$ statistics of the Gehrels' variance and is shown in Figure~\ref{fig:azmuth_beta} (\textit{left panel}). Like in other cooling flow clusters, X-ray emission from Abell 2390 also exhibit the central excess with respect to the 1D $\beta$-model and provides a strong evidence for the association of a cool core with this cluster. The $\beta$\,model is widely used in literature to investigate 2D distribution of ICM which yields the core radius and the slope of the surface brightness decrement. However, this 1d beta model resulted in a poor fit particularly in the central region. The resultant best fit parameters of the 1D $\beta$-model are: $r_c$ = 60\arcsec .10 ($\sim$ 217 kpc) and $\beta$ = 0.82 and are in 1$\sigma$ agreement with those obtained by \cite{Hicks06}. 

To account for the excess emission from the cool core of Abell 2390 we added another $\beta$-component to it. The resultant model is called as \textit{double-$\beta$ model} or \textit{two $\beta$-model} and takes the form :

\[\Sigma(r)=\Sigma_{01}\left[1+\left(\frac{r}{r_{c1}}\right)^2\right]^{-3\beta_{1}+0.5}\ + \Sigma_{02}\left[1+\left(\frac{r}{r_{c2}} \right)^2\right]^{-3\beta_{2}+0.5}\]

where the new surface brightness, core radius and slope-parameter are represented by $\Sigma_{01}$,  $r_{c1}$, and $\beta_{1}$,  respectively. Conventionally, the core radius of the central X-ray emission is presented by $r_{c1}$, while the overall X-ray emission of the cluster is presented by $r_{c2}$. Addition of this second $\beta$-component provided a better description for the X-ray emission from this cluster over the entire region with the best fit values $\beta_{2}$ = 0.74, $r_{c2}=19\arcsec.04$ ($\sim$ 68.92 kpc) and a steep component with slope parameter $\beta_{1}$= 10 to constrain the cusp part in the surface brightness distribution of Abell 2390. All fitting parameters given in Table~\ref{beta2d_t}.

To highlight fluctuations in the X-ray surface brightness distribution due to the presence of X-ray cavities and other features, we computed surface brightness profiles of the background-subtracted X-ray image of Abell 2390 by extracting 0.5$-$3.0\,keV counts from the four different sectors as marked in Figure~\ref{fig:unsharp}. Here, N-sector covers region of 20$^{\circ}$ -  130$^{\circ}$, E-sector covers 130$^{\circ}$ - 190$^{\circ}$, S-sector covers 190$^{\circ}$ - 290$^{\circ}$, while the W-sector covers 290$^{\circ}$ - 20$^{\circ}$, all angles measured in counter-clockwise direction. The resulting surface brightness profiles along with the best fit double $\beta$-model are shown in Figure~\ref{fig:azmuth_beta} ({\it right panel}). Dips in the surface brightness distributions at the locations of the X-ray cavities are clearly apparent in contrast to the best fitted double $\beta$-model. The deficit in the X-ray emission evident at $\sim$ 10\arcsec\,in the surface brightness profile along the W-sector hint towards the presence of cavities. The large depression apparent between 25 and 75\arcsec\, along the E Sector point towards the bigger deficit region along this direction.

\subsection{X-ray Cavity Detection}

\begin{figure*}
\vbox
{
\includegraphics[width=90mm,height=90mm]{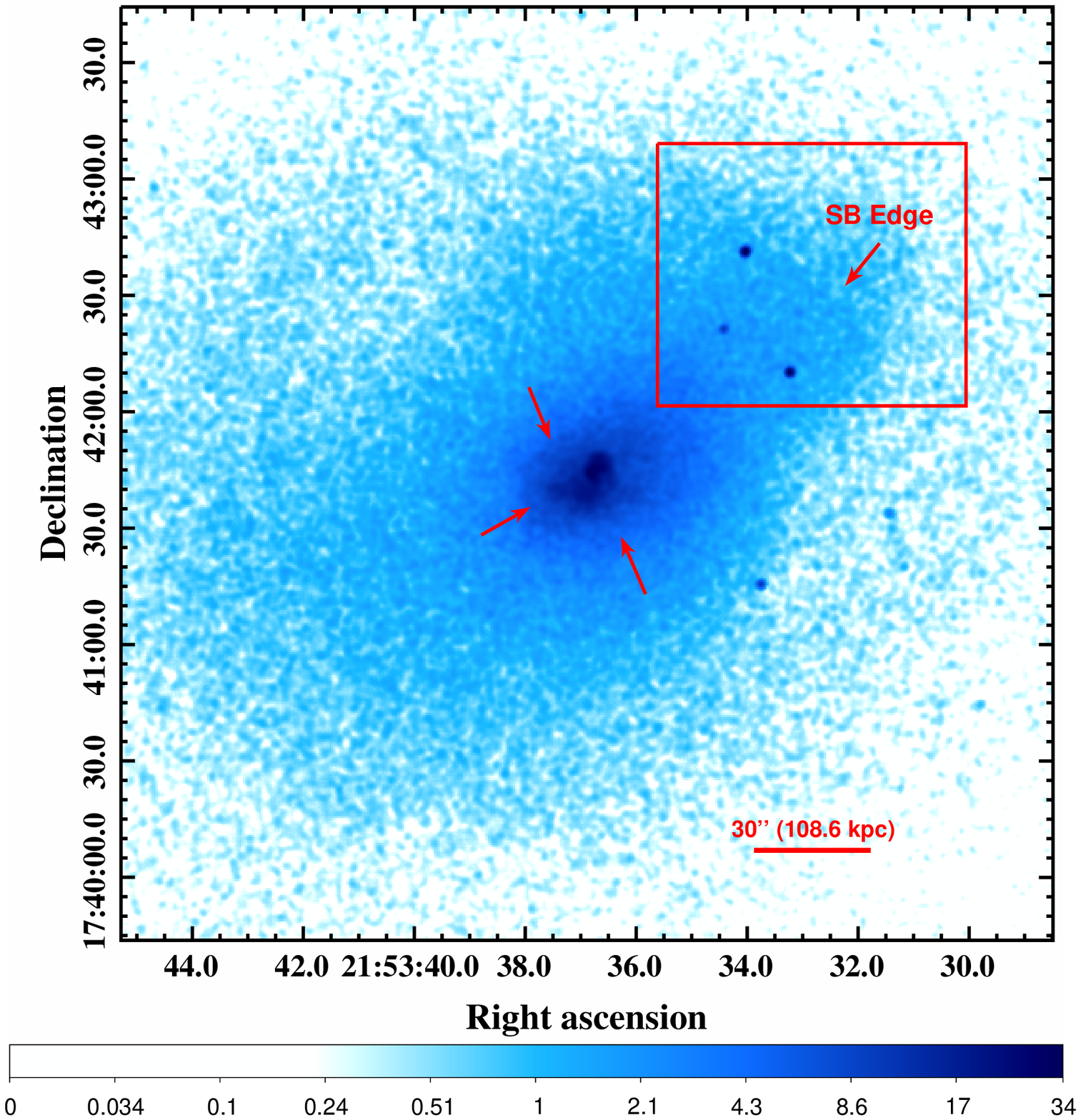}
\includegraphics[width=90mm,height=80mm]{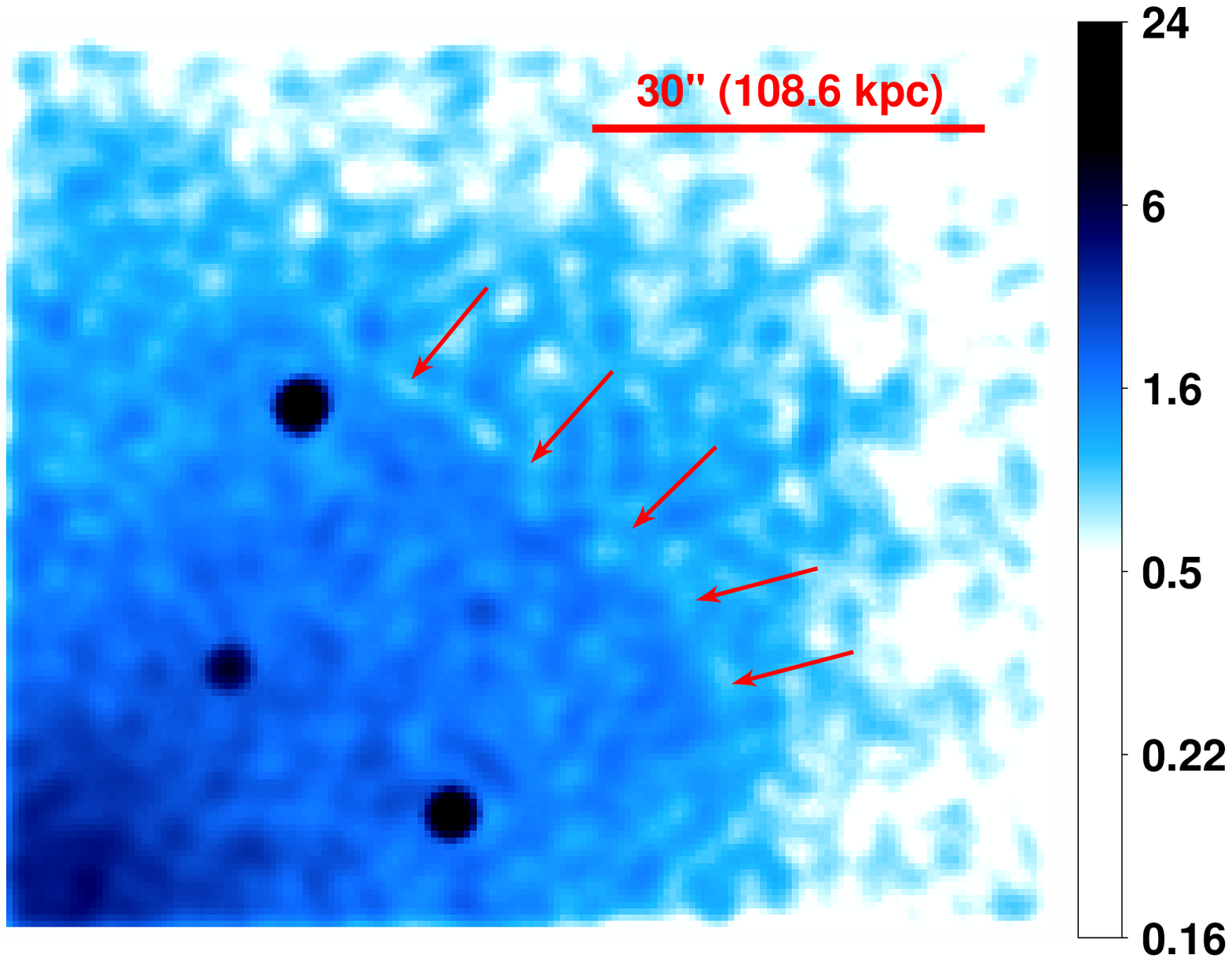}

}
\caption{{(\it left panel):} 0.3-7\,keV {\it Chandra} X-ray image of Abell 2390 in units of counts/pixel. This image has been smoothed with a 3 pixel (1.\arcsec5) Gaussian kernel and reveals asymmetric and elongated nature of the diffuse X-ray emitting gas. Red arrows near the centre shows region of temperature jump. An edge in the surface brightness has also been noticed in the north-western direction and is highlighted in the zoomed version of the figure ({\it right panel}).}
\label{fig:raw}
\end{figure*}

\begin{figure*}
{
\includegraphics[width=90mm,height=90mm]{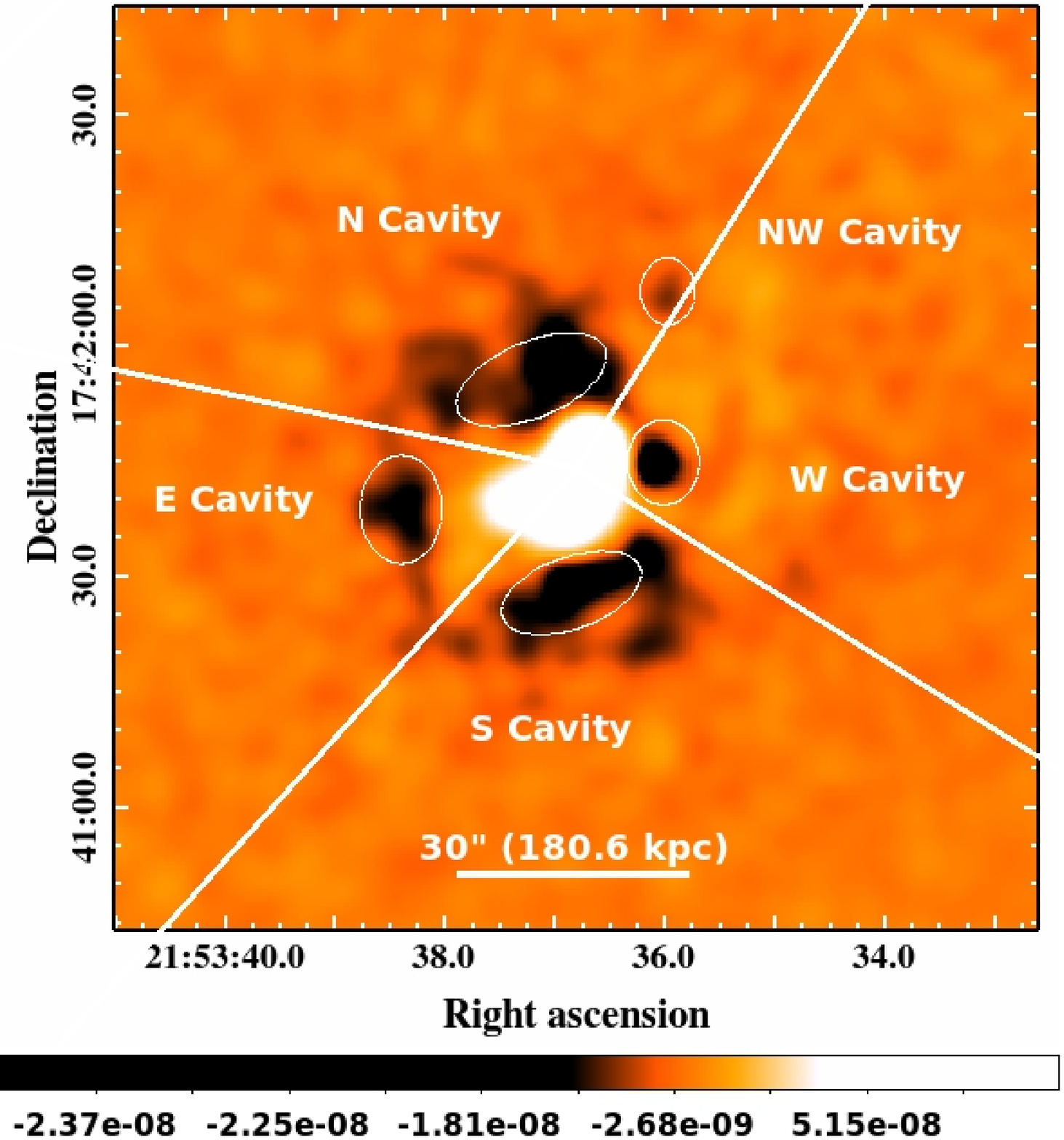}
}
\caption{ 0.5$-$3\,keV \textit{Chandra} unsharp-masked image of Abell 2390 in units of counts sec$^{-1}$ produced after subtracting a 12$\sigma$ wide Gaussian smoothed image from that smoothed with the 2$\sigma$ Gaussian. This figure confirms five X-ray deficient cavity regions (marked by white ellipses) in the surface brightness distribution in addition to the central excess emission. We also show the four different sectors N (20$\degr$-130$\degr$), E (130$\degr$-190$\degr$), S (190$\degr$-290$\degr$), and W (290$\degr$-20$\degr$), all angles measured in anti-clockwise manner, used for deriving the surface brightness profiles as well as temperature and metallicity profiles. The intensity representation in this figure has been reversed with respect to Figure~\ref{fig:raw} i.e., the darker shades in this figure represent the low  surface brightness.}
\label{fig:unsharp}
\end{figure*}

Though the presence of X-ray cavities and other features in Abell 2390 have been reported \citep{Vikhlinin05, Vikhlinin06, Russell13}, detailed analysis of the X-ray cavities using the latest 95 ks \textit{Chandra} data were not available in the literature. Therefore, in this paper we present systematic analysis of the latest dataset. Figure~\ref{fig:raw} (\textit{left panel}) clearly reveals  asymmetric distribution of the X-ray emission with an elongation along the north-west (NW) to the south-east (SE) direction and also hints towards the presence of a pair of X-ray cavities, one each on the north-east (NE) and the south-west (SW) direction of the X-ray peak. We investigate these structures in further details by employing a variety of image processing techniques such as, unsharp-masking, residual maps, IRAF\footnote{\color{blue}{{http://iraf.noao.edu/}}} ellipse fitting as well as by deriving the surface brightness profiles of the diffuse X-ray emission in different regions.

To identify the depressions in the \textit{Chandra} X-ray image, we created unsharp-masked image of Abell 2390 in the 0.5-3\,keV energy range by subtracting a strongly smoothed image ($\sigma =$12 pixels) from that smoothed lightly ($\sigma =$2 pixels) and is shown in Figure~\ref{fig:unsharp}. This figure reveals two prominent X-ray depressions 
(cavities) one each on the north-east and the other on south-west of the X-ray peak of the cluster centre. In addition to this, this figure also reveals three more relatively fainter X-ray cavities. These cavities in the unsharp masked image are highlighted by ellipses. 

\begin{figure*}
\vbox
{
\includegraphics[width=85mm,height=85mm]{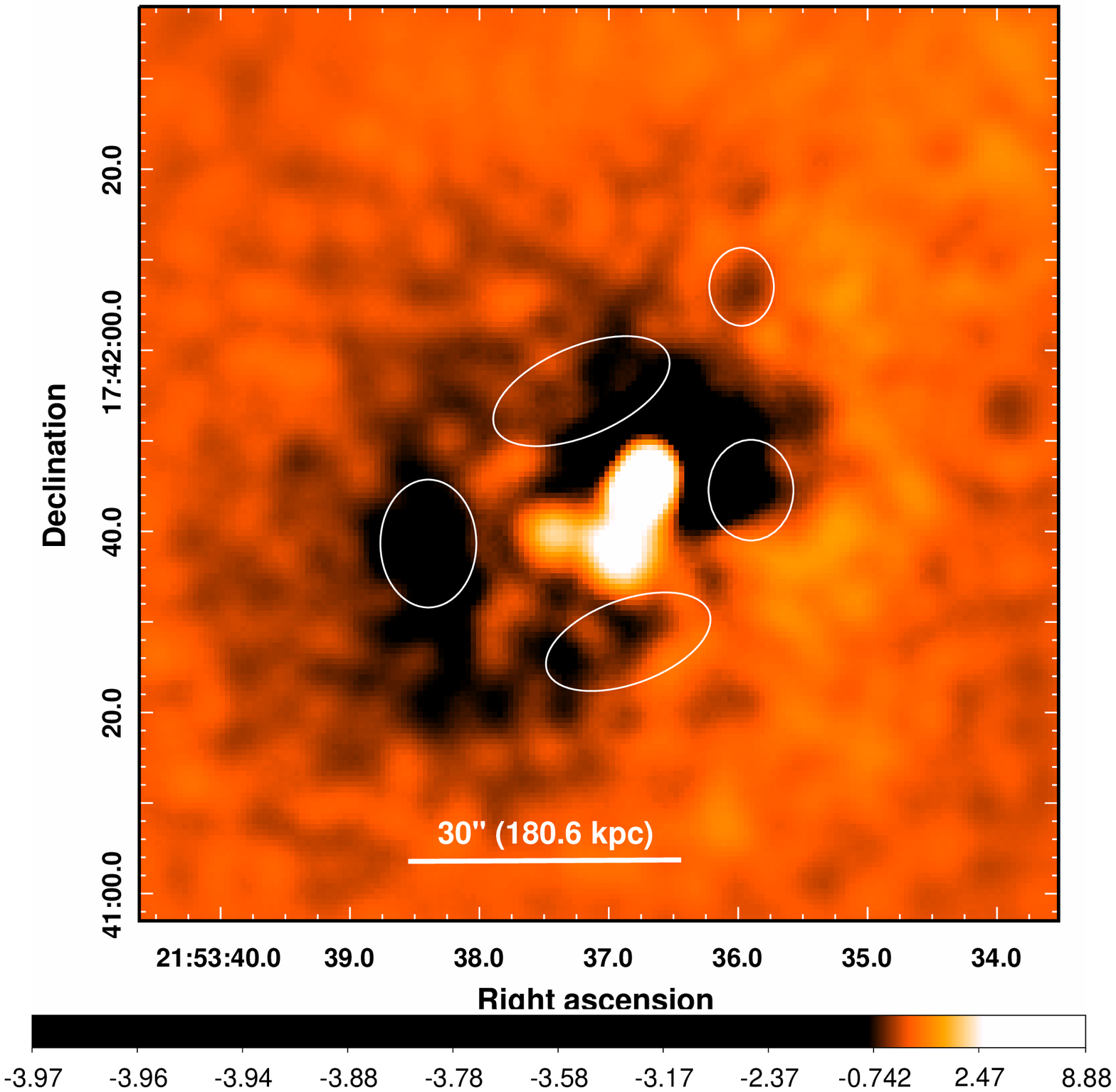}
\includegraphics[width=85mm,height=85mm]{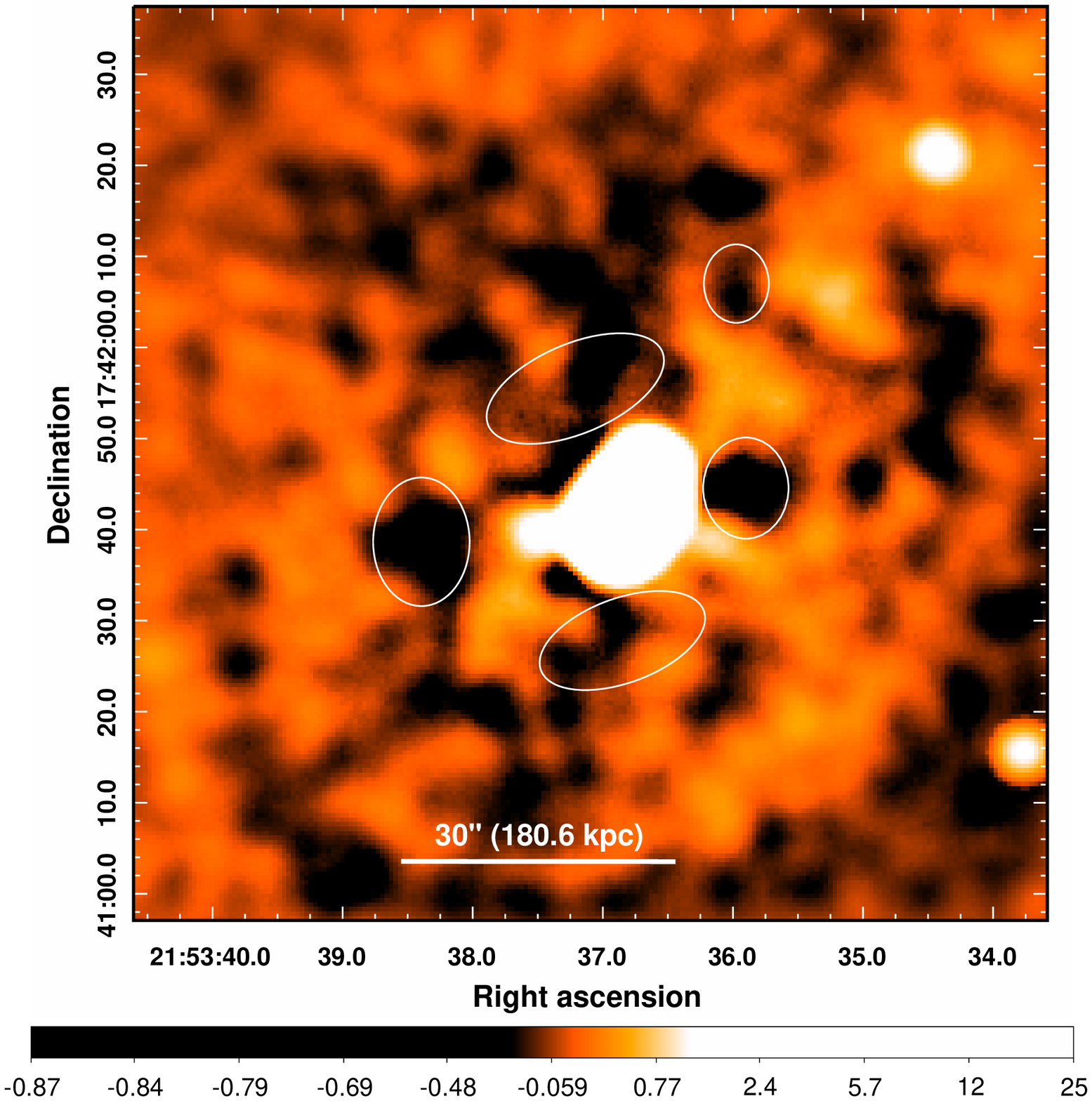}
}
\caption{{(\it left panel)}: 2-d elliptical $\beta$-model subtracted 0.5-3\,keV residual image of Abell 2390. This image has been smoothed with a 2\arcsec\,Gaussian kernel. This figure confirms the presence of X-ray cavities apparent in the unsharp-masked image of Abell 2390. The locations of X-ray cavities are highlighted by white ellipses. In addition to the central excess emission, an excess X-ray emission is also evident on the north-west direction, while deficiency in emission is apparent in cavity positions. {(\it right panel)}: ellipse fitted, 0.5-8\,keV \textit{Chandra} residual image using IRAF task \texttt{ellipse} and further smoothed with 3\arcsec Gaussian.}
\label{fig:beta2d}
\end{figure*}

To better visualize the structures in the surface brightness distribution in two dimensions, we created residual map of Abell 2390 by subtracting the simulated smooth elliptical 2d $\beta$ model from the cluster emission generated fitting a two-component 2d $\beta$ model to the surface brightness distribution of X-ray emitting gas. The resultant residual map is shown in Figure~\ref{fig:beta2d} ({\it left panel}), which reveals the structures apparent in the unsharp masked image. However, as this 2d $\beta$ model assumes spherical symmetry as well as hydrostatic equilibrium of the of the gas, therefore the results at smaller radii do not match perfectly with those seen in the unsharp masked image. Particularly, the morphologies of the X-ray cavities in the two images are not identical. However, the actual distribution of the X-ray emitting gas exhibits a very complex structure and vary strongly as a function of position angle. As a result the output image do not represent the true distribution of the gas and hence the morphological difference between the residual map and the unsharp masking image. 
The excess X-ray emission (bright shades) in the central region as were seen in the unsharp mask image are also apparent in this figure. We also checked morphology as well as presence of the cavities and other features by subtracting elliptical model by fitting ellipses to the iso-intensity contours in the cleaned X-ray image using the task {\texttt ellipse} available within IRAF. While fitting the ellipses, centre coordinates were fixed at the location of the BCG, whereas the eccentricity and position angle were allowed to vary. Resultant elliptical model subtracted residual map is shown in Figure~\ref{fig:beta2d} ({\it right panel}).

\begin{figure*}
\vbox
{
\includegraphics[width=90mm,height=90mm]{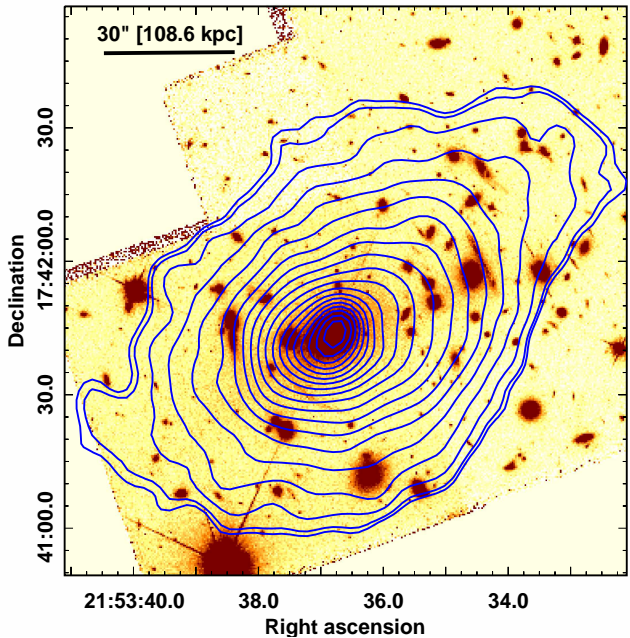}
}
\caption{To examine extent of the X-ray diffuse emission from the cluster we overlay the 0.3-7\,keV iso-intensity contours on the HST F813W band image of Abell 2390. Cluster diffuse emission is asymmetric and elongated along the north-western to the south-eastern direction. The X-ray peak lies within 0.5 arcsec of the optical centre of the cD galaxy PGC 140982. }
\label{fig:rawoptical}
\end{figure*}

To compare the extent of the X-ray emission from Abell 2390 with its optical counterpart, we overlay the 0.3-7\,keV X-ray iso-intensity contours on the HST I band (F814W) image (Figure~\ref{fig:rawoptical}). This figure clearly reveals the asymmetric and elongated nature of the X-ray emission with its centre coinciding with the cD galaxy PGC 140982, whose optical centre lies within 0.\arcsec 5 of the X-ray centre of Abell 2390.  

\subsection{X-ray Surface Brightness Edge}

The {\textit{Chandra} raw image in the energy 0.3-7\,keV, smoothed with 3 pixel Gaussian kernal shown in Figure~\ref{fig:raw}, revealed X-ray surface brightness (SB) edge at $\sim$68 arcsec
on NW direction of the X-ray centre of Abell 2390 (highlighted by red arrows). The zoomed in view of the surface brightness edge is shown in Figure~\ref{fig:raw} ({\it right panel}).
This SB edge appears like $\sim$110\,kpc long X-ray arc. Deeper X-ray observations of this cluster are called for to explore this SB edge in further detail. This X-ray surface brightness edge is found to coincide with the complex edge in the radio diffuse emission map obtained using 1.4\,GHz VLA calibrated images available in the archive of NRAO \footnote{\color{blue}{{https://archive.nrao.edu/archive/}}}.

\subsection{Substructures in the central region}
\begin{figure*}
{
\includegraphics[width=70mm,height=70mm]{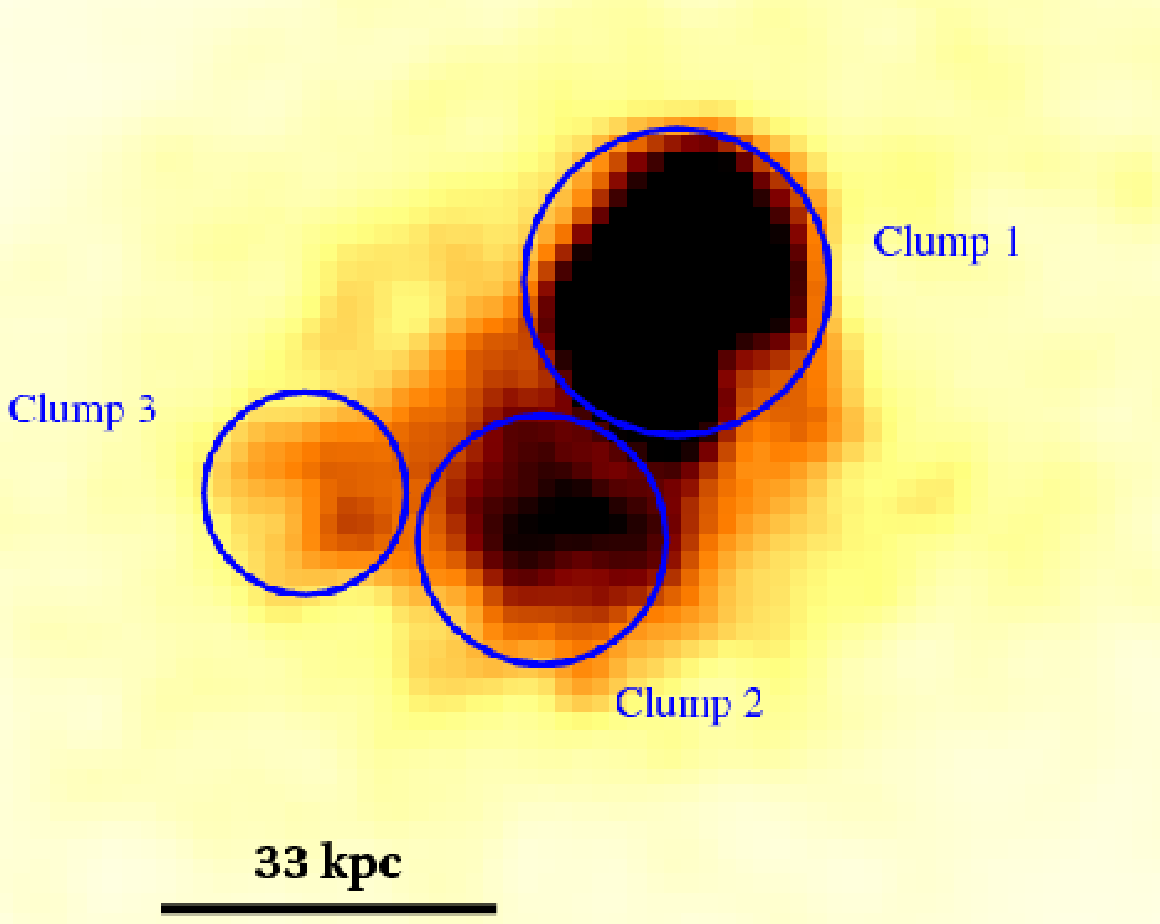}
\includegraphics[width=70mm,height=70mm]{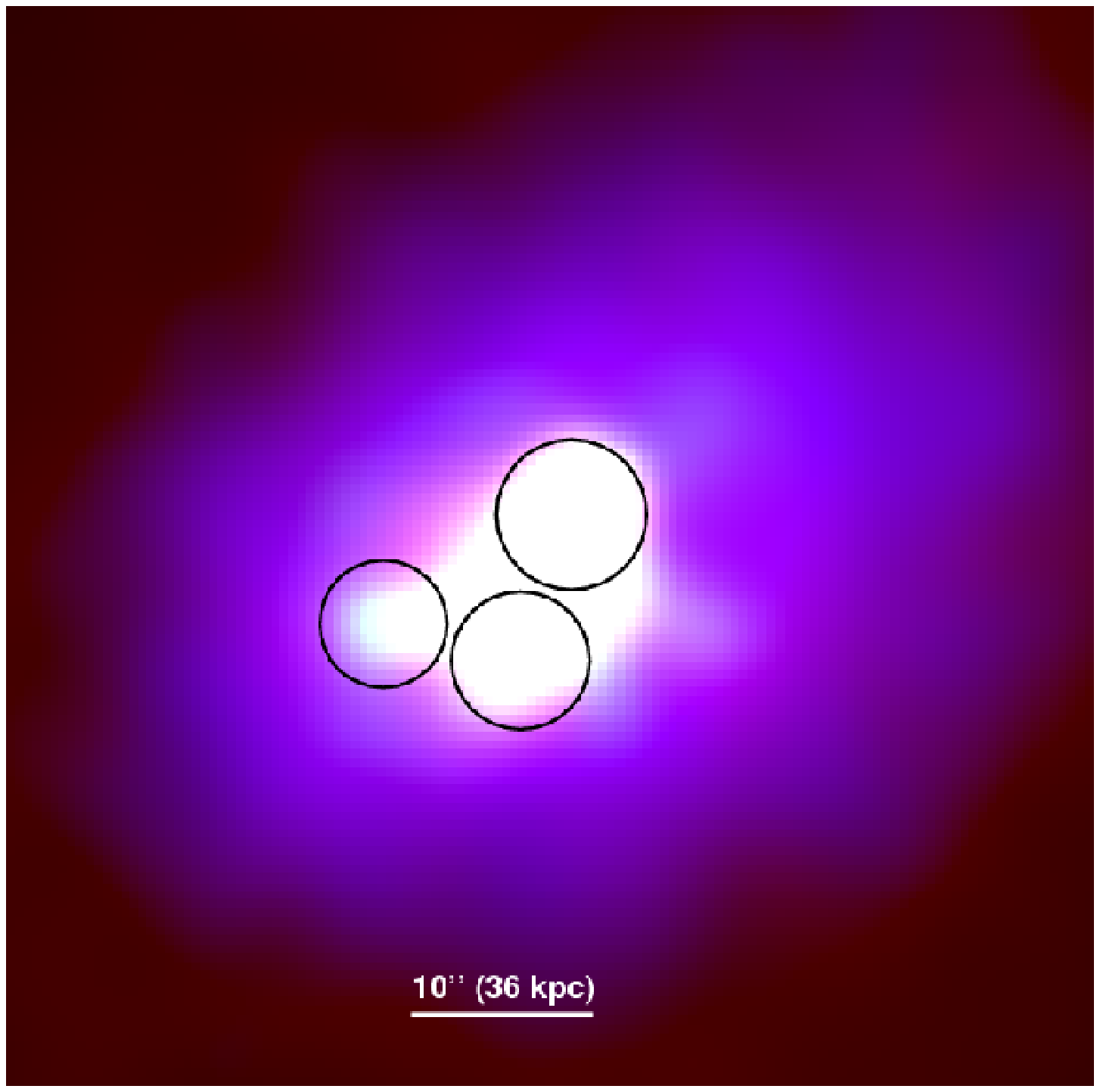}
}
\caption{{(\it left panel):} 0.5-3\,keV exposure corrected background subtracted 3$\sigma$ smoothed \textit{Chandra} image of the central 1\arcmin  region of Abell 2390. This figure clearly revels different clumpy knot-like features. Spectra representative from each of the clump marked in the figure have been extracted separately for investigating their thermodynamical properties. {(\it right panel):} RGB tri-colour image of the central 1 arcmin of Abell 2390 derived after combining X-ray photons in three different energy bins, namely, soft (0.3 - 1\,keV, shown in red colour), medium (1 - 2\,keV, shown in green), and hard band (2 - 10\,keV, shown in blue). Apparent positions of the clumps are marked by black circles. }
\label{fig:clumps}
\end{figure*}

0.5-3\,keV \textit{Chandra} blank-sky background subtracted exposure corrected image created for the central 1\arcmin\,($\sim$ 217 kpc) region reveals clumpy or knot-like complex substructures 
(Figure~\ref{fig:clumps} {\it left panel}) similar to that reported earlier in several other systems, e.g. Centaurus A \citep{Sanders01}, Abell 1991 \citep{Sha04, Pan13}, 
2A0335+096 \citep{Mazz03}. These clumpy sub-structures are comprised of several finer knot-like structures on arcsecond scales and are connected to one another.  

To examine structure of these clumpy regions we produced a tri-colour map of the central region of Abell 2390. For this, we filtered the \textit{Chandra} image in three different energy bands namely, the soft (0.5-1\,keV, shown in red colour), the medium (1-2\,keV, shown in green) and the hard band (2-10\,keV, shown in blue) and were then combined after proper scaling. The adaptively smoothed tri-colour map of the central 1 arcmin ($\sim$ 217 kpc ) of Abell 2390 is shown in Figure~\ref{fig:clumps} {(\it right panel)}, which illustrates the clump-like structures in the medium band. This figure reveals that the central substructures are due to the relatively warmer components compared to its surrounding.

\section{Spectral Analysis}
\label{spec}
To examine spectral properties of hot gas in the regions of enhanced and suppressed X-ray emission, we extracted 
X-ray spectra from each of the region of interest separately using the CIAO script \texttt{specextract}. 
For each extraction, background spectra were obtained from the same region in the ``blank-sky" and were normalized 
equating the 9-12\,keV count rates of the observed and background data. Weighted responses were generated for each 
extraction and the spectral fitting was performed using the {\sc XSPEC} version 12.7.1 \citep{Arnaud96}. 

\begin{table*}
\caption{Summary of the spectral analysis of the gas extracted from
different regions of interest}
\begin{tabular}{@{}cccccccr@{}}
\hline
\hline
{\it Regions}  &{\it Best-fit} & $kT$ & Abundance &$\chi^{2}$/dof & Net Counts  
\\
 &{\it model} &(keV) & ($Z_{\odot}$)&  \\
\hline
\hline
\textit{N - cavity}  & (PHABS*APEC) &$7.76_{-0.61}^{+0.60}$
&$0.48_{-0.14}^{+0.15}$ &135.66/150 = 0.90 &3478  \\
\textit{S - cavity}  & (PHABS*APEC) &$7.67_{-0.68}^{+0.69}$
&$0.26_{-0.12}^{+0.13}$ &101.21/134 = 0.75 &2950  \\
\textit{E - cavity}  & (PHABS*APEC) &$9.63_{-1.36}^{+1.76}$
&$0.83_{-0.35}^{+0.37}$ &81.48/78 = 1.04 &1510  \\
\textit{W - cavity}  & (PHABS*APEC) &$7.42_{-0.70}^{+0.74}$
&$0.51_{-0.16}^{+0.17}$ &106.4/120 = 0.88 &2467  \\
\textit{NW - cavity}  & (PHABS*APEC) &$8.50_{-1.90}^{+4.01}$
&$1.13_{-0.92}^{+1.02}$ &13.67/18 = 0.91 &376  \\
\textit{Gas Clump 1 (4.\arcsec17)}  & (PHABS*APEC) &$3.70_{-0.14}^{+0.14}$
&$0.50_{-0.07}^{+0.08}$ &194.84/162 = 1.20  &5038 \\
\textit{Gas Clump 2 (3.\arcsec40)}  & (PHABS*APEC) &$4.88_{-0.25}^{+0.27}$
&$0.74_{-0.13}^{+0.14}$ &143.56/134 = 1.07  &3253 \\
\textit{Gas Clump 3 (2.\arcsec77)}  & (PHABS*APEC) &$5.21_{-0.37}^{+0.44}$
&$0.42_{-0.14}^{+0.15}$ &45.07/62 = 0.73  &2058 \\
\textit{Total Gas Clumps} & (PHABS*APEC) &$4.45_{-0.10}^{+0.16}$
&$0.53_{-0.06}^{+0.06}$ &224.06/224 = 1.00  &10375 \\
\textit{SB Edge} & (PHABS*APEC) &$7.13_{-0.46}^{+0.49}$
&$0.42_{-0.09}^{+0.10}$ &179.08/195 = 0.92  &5826 \\
\textit{Cluster Diffuse Emission}  &(PHABS*APEC) &$8.27_{-0.09}^{+0.09}$ &
$0.33_{-0.02}^{+0.02}$ &527.54/453 = 1.17 & 178445 \\

\hline
\hline
\end{tabular}
\footnotesize 
\begin{flushleft} 
{Notes:} col 1 - regions of interest explored by spectral analysis 
in the energy range 0.5$-$7\,keV (quantities in the bracket indicate radius), 
col 2 - best-fit model, col 3 \& 4 - temperature and abundance, 
respectively, col 5 - goodness of the fit, and col 6 - net  
X-ray counts extracted from the regions of interest.
\end{flushleft}
\label{spectralpro}
\end{table*}

\begin{figure*}
{
\includegraphics[width=80mm,height=80mm]{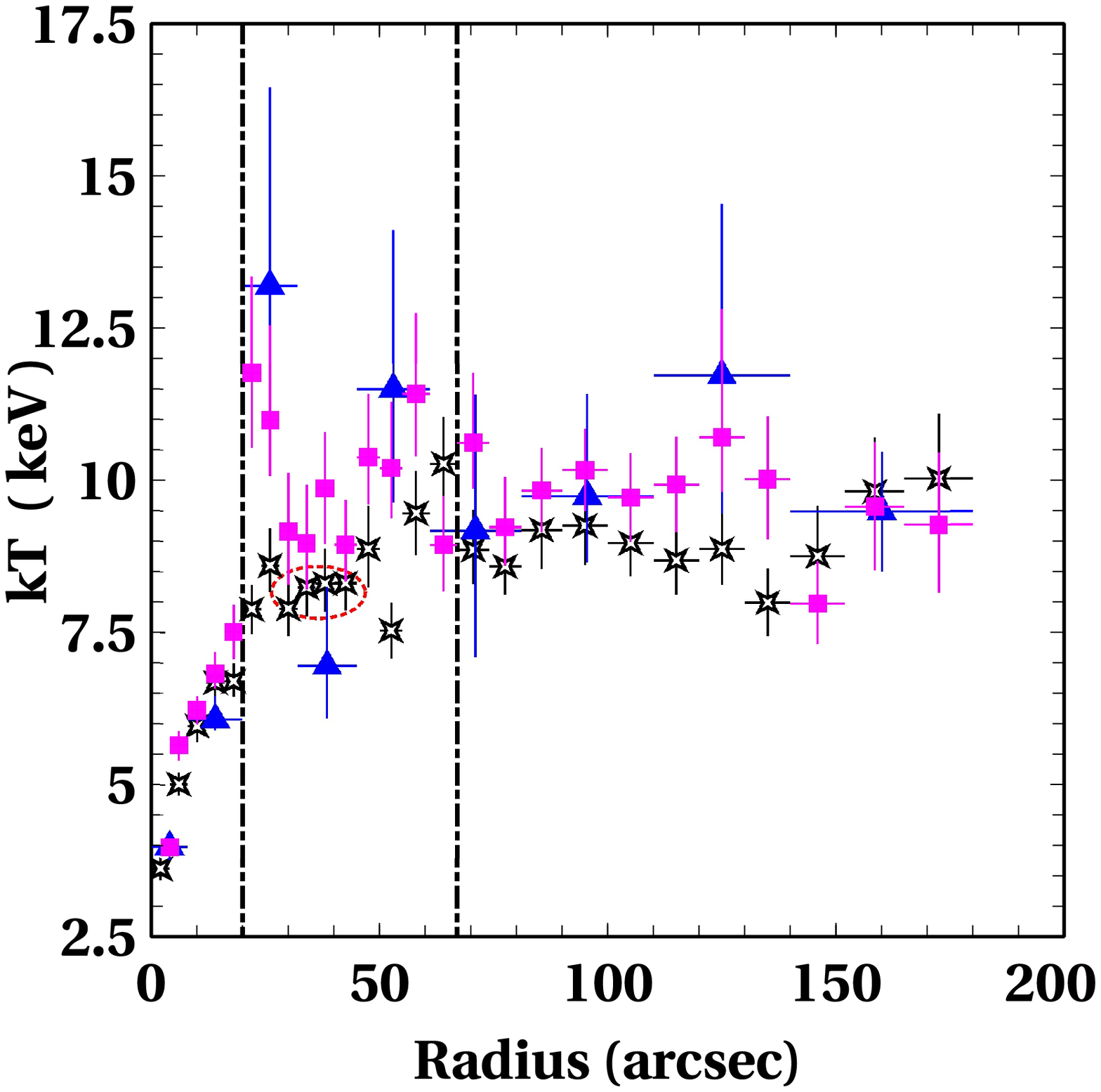}
\includegraphics[width=80mm,height=80mm]{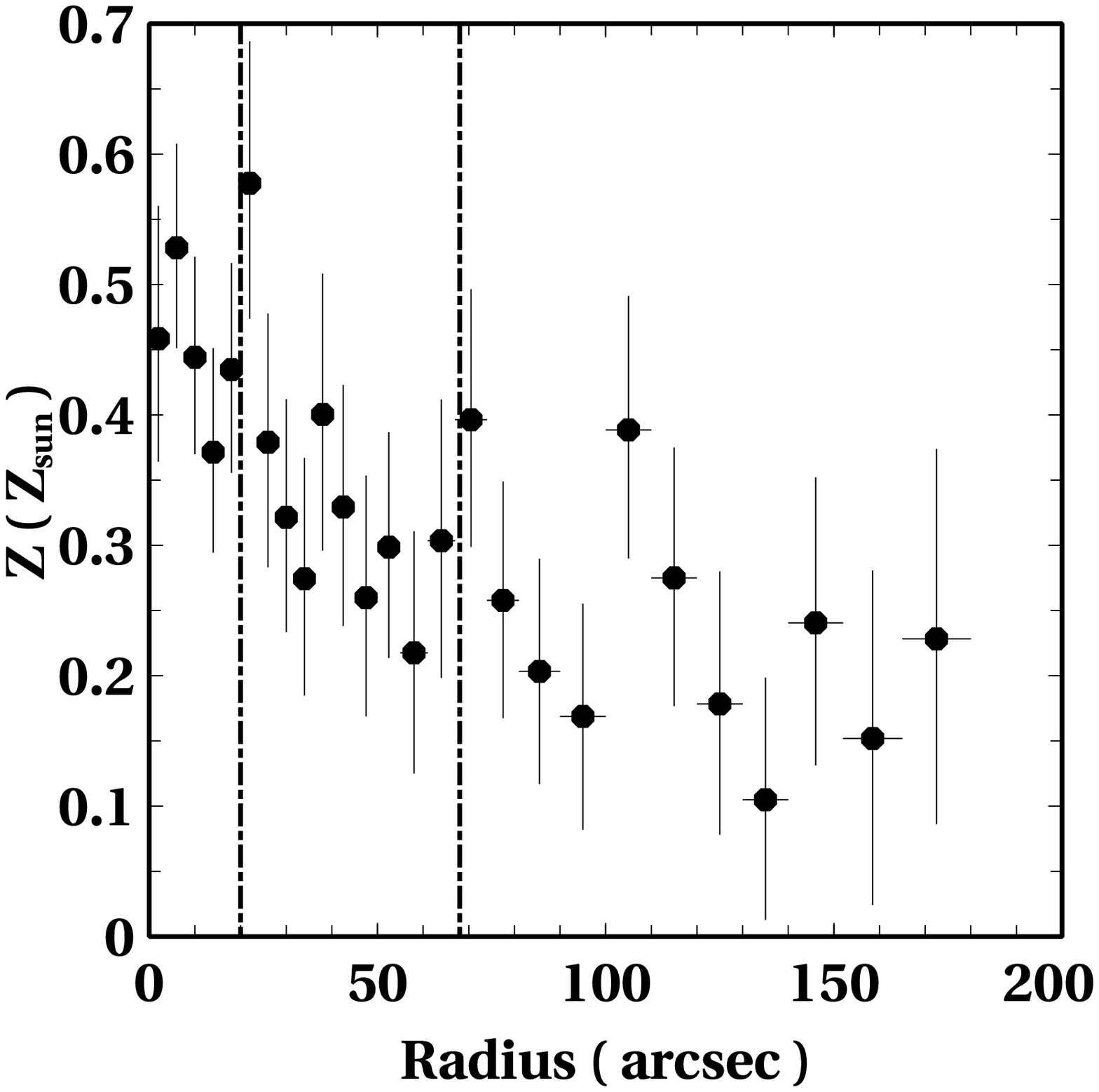}
\includegraphics[width=80mm,height=80mm]{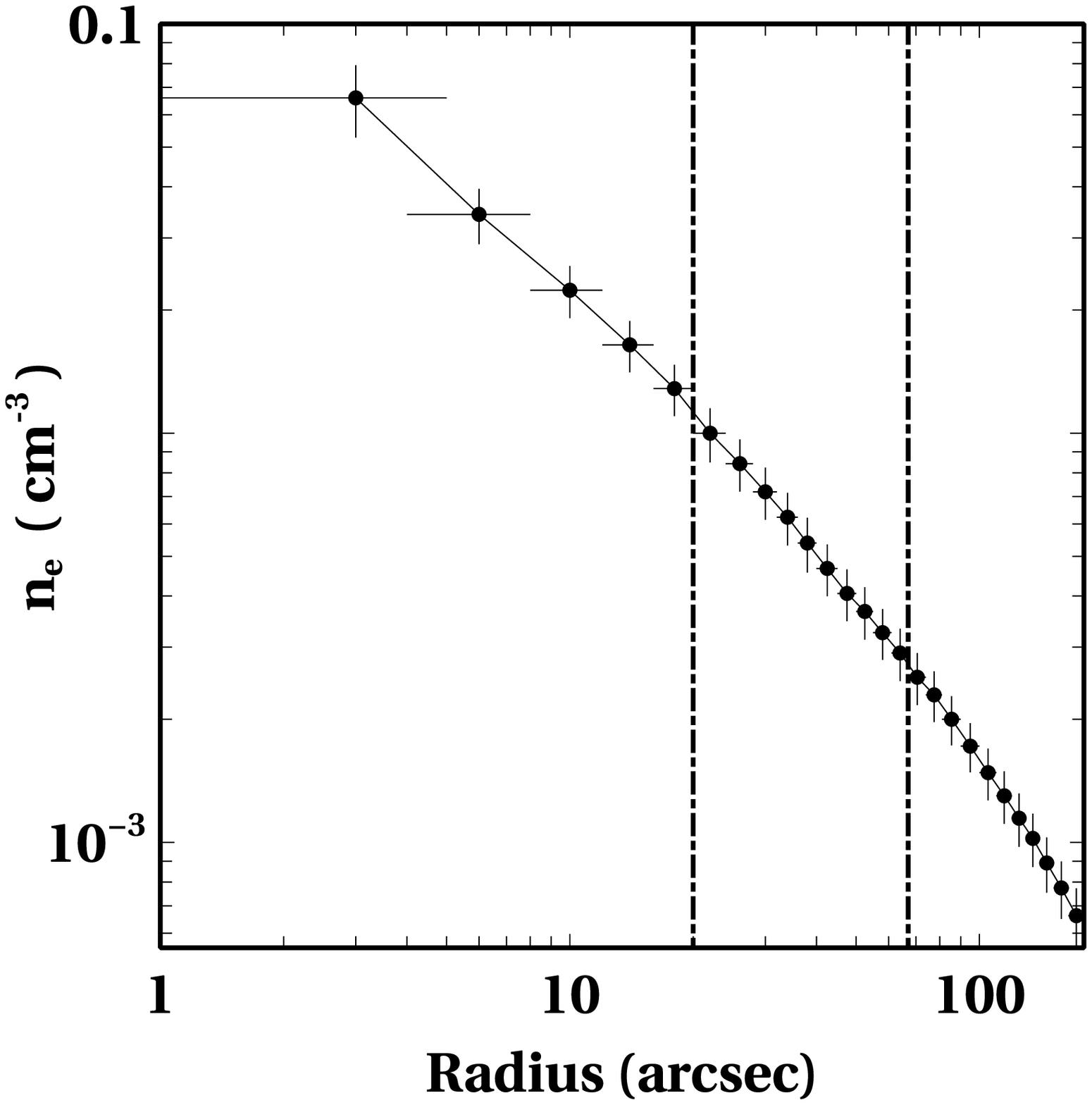}
\hspace{0.4cm}
\includegraphics[width=80mm,height=80mm]{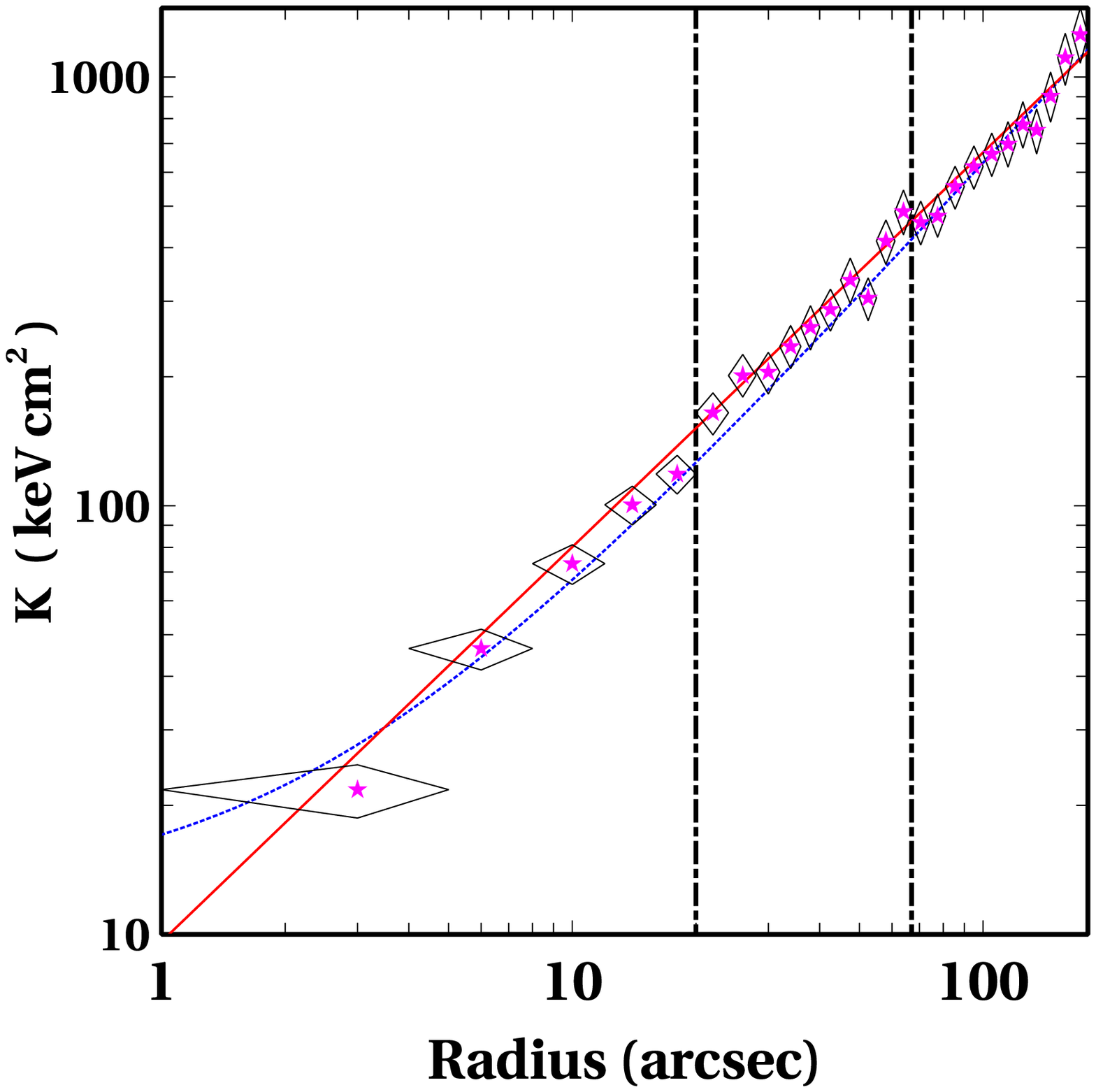}
}
\vspace{-23pt}
\caption{{\it Top left}: Projected radial temperature profile of the X-ray emission extracted from elliptical annuli centred on the cluster centroid (open stars). The filled magenta squares indicates the projected temperature variation after excluding the cool region from the cluster environment as seen in the hardness map (Figure~\ref{fig:KT1_cool}). In the same figure we also plot the temperature profile of the X-ray emission from deprojection analysis (shown in filled blue triangles).  The metallicity {(\it top right)}, electron density {(\it bottom left)} and entropy profiles {(\it bottom right)} of the diffuse X-ray emission from Abell 2390 are plotted as a function of the 
radial distance. The dotted vertical lines in these profiles represent the locations of temperature jumps.}
\label{fig:KT1_Z1}
\end{figure*}

\subsection{Profiles of temperature and other parameters}
To examine the radial dependence of the gas properties, we computed projected radial temperature and metallicity profiles by fitting 0.5$-$7\,keV spectra extracted from elliptical annuli centred on the centroid of the diffuse emission from Abell 2390.
Each elliptical annulus was taken to have the major-to-minor axis ratio and position angle as set above and the radial bin width of each annulus was adjusted so as to get at least $\sim$ 6,400 background subtracted counts (S/N$\sim$ 80). Source spectra, background spectra, photon-weighted response files and photon-weighted effective area files were generated for each of the elliptical annulus using the CIAO task \texttt{specextract} and was fitted with an absorbed single-temperature {\sc apec} model within XSPEC adopting the $\chi^2$-statistics. Temperature, abundance and normalization parameters were allowed to vary during the fit, while the foreground column density was fixed at the Galactic value of $N_H = 1.07 \times 10^{21}$ cm$^{-2}$ \citep{Vikhlinin05}.

\begin{figure*}
\vbox
{
\includegraphics[width=170mm,height=170mm]{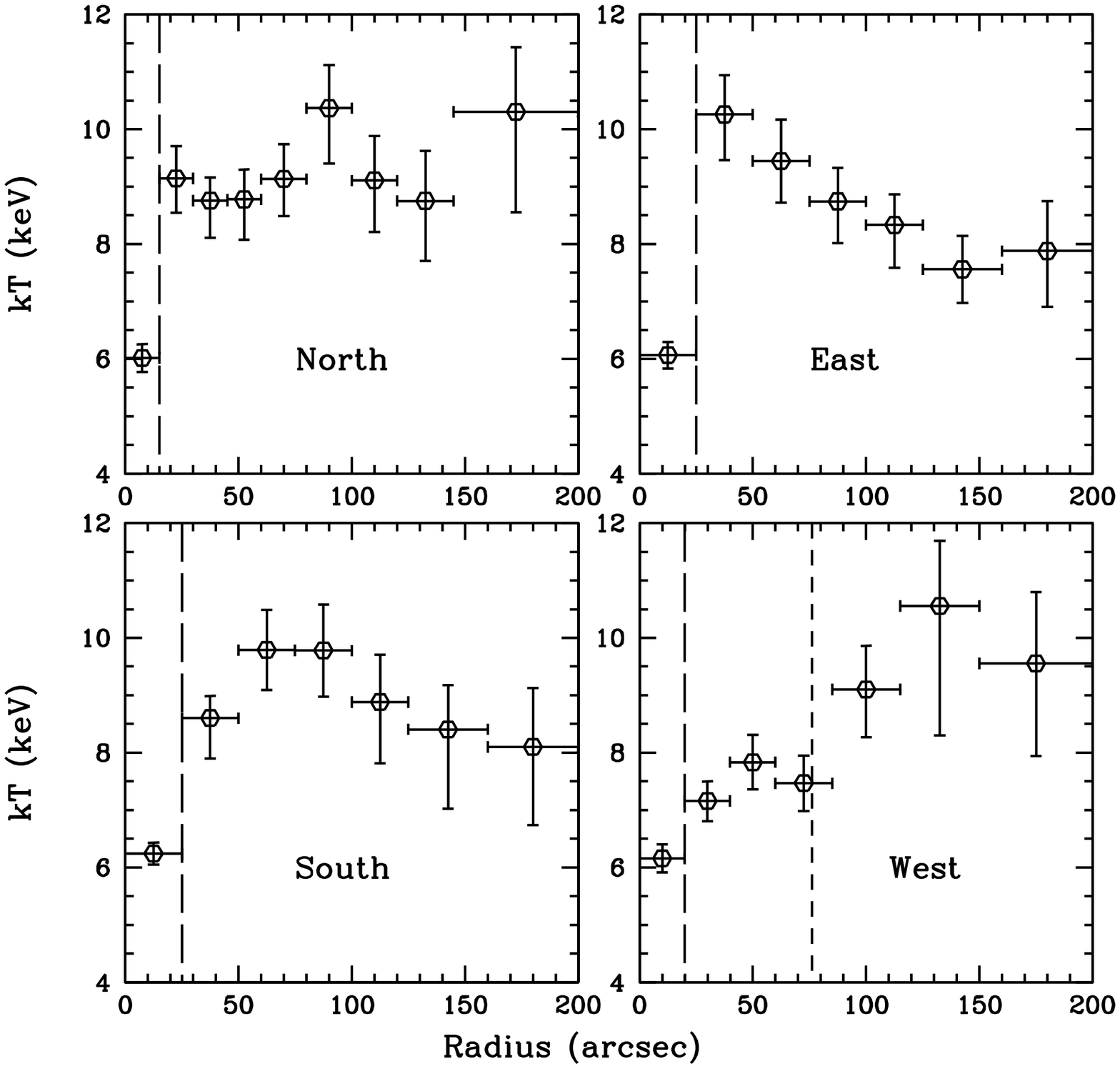}
}
\caption{Projected radial temperature profiles computed along four different (N, E, S and W) wedge shaped sectors (Figure~\ref{fig:unsharp}). Widths of the annuli in each sector were adjusted such as to achieve a S/N $\sim$80 (6400 counts). The vertical dashed lines represent the locations of temperature jumps.}
\label{fig:temp_prof}
\end{figure*}

Resulting projected temperature profile derived from this analysis is shown in Figure~\ref{fig:KT1_Z1} {(\it top left)} by open stars. This profile clearly reveals an increase in temperature of the X-ray emitting gas from a minimum of 6.69\,keV to 7.88\,keV at about 20\arcsec. Then it drops to a value of 7.5\,keV and remains constant upto 45\arcsec\,shown by red ellipse. It once again show a rise upto 10.5\,keV at 68\arcsec\,and again drops to 7.88\,keV, which latter remains roughly constant upto about 150\arcsec. Beyond 150\arcsec\,temperature again rises to 10.20\,keV. Prominent temperature jumps at 20\arcsec\,and 68\arcsec\,are highlighted by vertical dotted lines and are probably associated with cold fronts. Similar trends in the temperature profiles have been reported in clusters with cooling cores \citep{Gastaldello07, Rasmussen07, David09, Sun09, Gitti11}.

In order to investigate the origin of the constant temperature seen in the projected temperature profile from $\sim$ 30\arcsec\,to 45\arcsec\,highlighted by red ellipse. We created hardness ratio map by dividing a 1.5-8 keV image by a 0.3-1.5 keV image. Before division both images were smoothed by a Gaussian kernel of 10 pixel (5\arcsec\,). The resultant map is shown in Figure~\ref{fig:KT1_cool}. In the figure dark shades are represents regions of low temperature gas. These regions of cooler gas are extending towards the north-east and the south-west side of the cluster. By visual inspection of the hardness ratio map, we selected most prominent regions marked by blue regions (see Figure~\ref{fig:KT1_cool}). We then excluded these regions and extracted the spectra from the same annuli and generated the temperature profile shown by filled square data points in Figure~\ref{fig:KT1_Z1} ({\it top left}). As evident from the comparison of the two profiles, the stage of constant temperature has been removed and the temperature profile look like cool core clusters. This clearly indicates that the marked regions must contains low temperature gas. Cool gas excluded profile also confirm the temperature jump at 20\arcsec\,and at 68\arcsec\, which then remains roughly constant at 10 keV.

To determine the three-dimensional structure of the ICM we performed a deprojection analysis of the X-ray photons extracted from concentric elliptical annuli as above, but in this case widths of the elliptical annuli were adjusted so as to achieve S/N$\sim$40. Using the ``onion peeling" method of \citet{blan03}, we derive the deprojected temperature profile. Here, temperature, abundance and normalization parameters were first derived for the spectrum extracted from the outermost annulus by fitting it with an absorbed {\sc apec} model. Then the spectrum from annulus interior to it was fit by adding one more {\sc apec} component to that for the outer annulus, with fixed temperature and abundance but normalization scaled to project from the outer to the inner annulus and the procedure was continued till the centre of the cluster. Profile of the deprojected temperature is plotted in the same figure (Figure~\ref{fig:KT1_Z1} {\it top left}) and is shown by filled blue triangles. A comparison of the two temperature profiles reveal that the deprojected temperature profile roughly follow the same trend as of the projected analysis. Here a sharp temperature rise from 6\,keV to 13.25\,keV is seen at about 20\arcsec\,which then drops to 7\,keV indicating towards a cold front.

To investigate these temperature jumps in further details, we compute the temperature profiles using the spectral extraction from four different sectors (N, E, S, and W) as used above for computing the surface brightness profiles. Annular regions in each sector were selected so as to have a minimum of 6400 background subtracted counts (S/N$\sim$80). Extracted spectra were then treated in the same way as discussed above in the radial profiles and the results are shown in Figure~\ref{fig:temp_prof}. Temperature profile derived along the W-sector shows a monotonic rise from 6.16\,keV to 7.16\,keV at 20\arcsec, also temperature jump observed at the location of SB edge from 7.47\,keV to 9.10\,keV. Temperature profiles along the N sector show jump from 6\,keV to 9.25\,keV at $\sim$15\arcsec\,(54 kpc), along the S sector from 6.3\,keV to 8.5\,keV at $\sim$25\arcsec\,(90.5\,kpc), along the E sector 6\,keV to 10.27\,keV at $\sim$25\arcsec\,(90.5\,kpc) and along the W sector from 6.16\,keV to 7.16\,keV at $\sim$20\arcsec\,(72.4\,kpc). These jumps apparent in the temperature profile along the N, S, E and W sector provide the clearest indication regarding their association with the cold fronts. The same locations of jumps in all the profiles, indicates that cold fronts covers nearly circular region surrounds the BCG. After confirming jumps in the temperature profiles along the all sectors, we next compute the Mach number associated with the cold fronts. The Mach number corresponding to the jump along N sector is 1.54$\pm$0.08, S sector is 1.36$\pm$0.11, E sector is 1.69$\pm$0.09 and W sector is 1.17$\pm$0.10. We got average Mach number as 1.44$\pm$0.05. Whereas, Mach number at SB edge is 1.22$\pm$0.06.

The radial metal abundance profile computed for Abell 2390 is shown in {Figure~\ref{fig:KT1_Z1} ({\it top right}). The metal abundance at the location of temperature jumps (at about 20\arcsec\,and 68\arcsec\,) exhibit a small rise, and are consistent with that seen in other cool core clusters \citep{Allen98, Gastaldello07, Rasmussen07, David09, Sun09, Pan12, Pan13}.

At the cluster temperature computed above, the ICM is hot and nearly fully ionized. Therefore, the surface brightness distribution can be considered as a tracer of the electron number density $n_e$. Hence using $n_e = n_{e0} \left(1+(\frac{r}{r_c})^2\right)^{-3\beta /2}$, we compute the electron number density and its variation in the form of electron density profile which is shown in Figure~\ref{fig:KT1_Z1} ({\it bottom left}). For this we used the normalization parameter of the {\sc apec} component in radial temperature profile and assumed $n_e = 1.2\,n_H$, for a fully ionized gas with hydrogen and helium mass fractions of $X\,=\,0.7$ and $Y\,=\,0.28$, respectively. 

\subsection{Entropy Profile}

The entropy index of the ICM offers a direct probe to investigate heating and cooling processes in the central region of the clusters relative to the individual temperature and density profiles. Owing to which we also derive the entropy, defined as $K = kT n_e^{-2/3}$, profile of the X-ray emission distribution from Abell 2390 and is shown in  Figure~\ref{fig:KT1_Z1} (\textit{bottom right}). The radial entropy profile derived for Abell 2390 is found to fall systematically in the radially inward direction. To examine its nature we fit this entropy profile by a powerlaw model $K(r) = K_{0} + K_{100} (r/100\,kpc)^{\alpha}$ \citep{Cav09}. Here, $K_{0}$ represents the core excess entropy or entropy floor above the best-fit powerlaw, $K_{100}$ the entropy at 100 kpc, and $\alpha$ the power law index ($\sim$1.07). The best fit model is shown by the blue dotted line in this figure with $K_{100}$ = $204\pm 22.84$\,keV $cm^{2}$ and the entropy floor of $K_{0}$ = 12.20$\pm$2.54\,keV$cm^{2}$. Thus, flattening of the mean entropy index towards the central projected value of 12.20$\pm$2.54\,keV$cm^{2}$ indicate towards an intermittent heating of the ICM. If cooling of the ICM is not compensated by heating, the ICM will radiate away whole of its thermal energy on a timescale of $t_{cool} \le$ 1.9 Gyr \citep{hla11, Cav09} and the entropy profile will follow the nature shown by red continuous line in this figure which represent a pure cooling system. Notice the significant deviation in the central region pointing towards the entropy floor at $12.20\pm 2.54$\,keV $cm^{2}$.

\begin{figure*}
{
\centering
\includegraphics[scale=0.8]{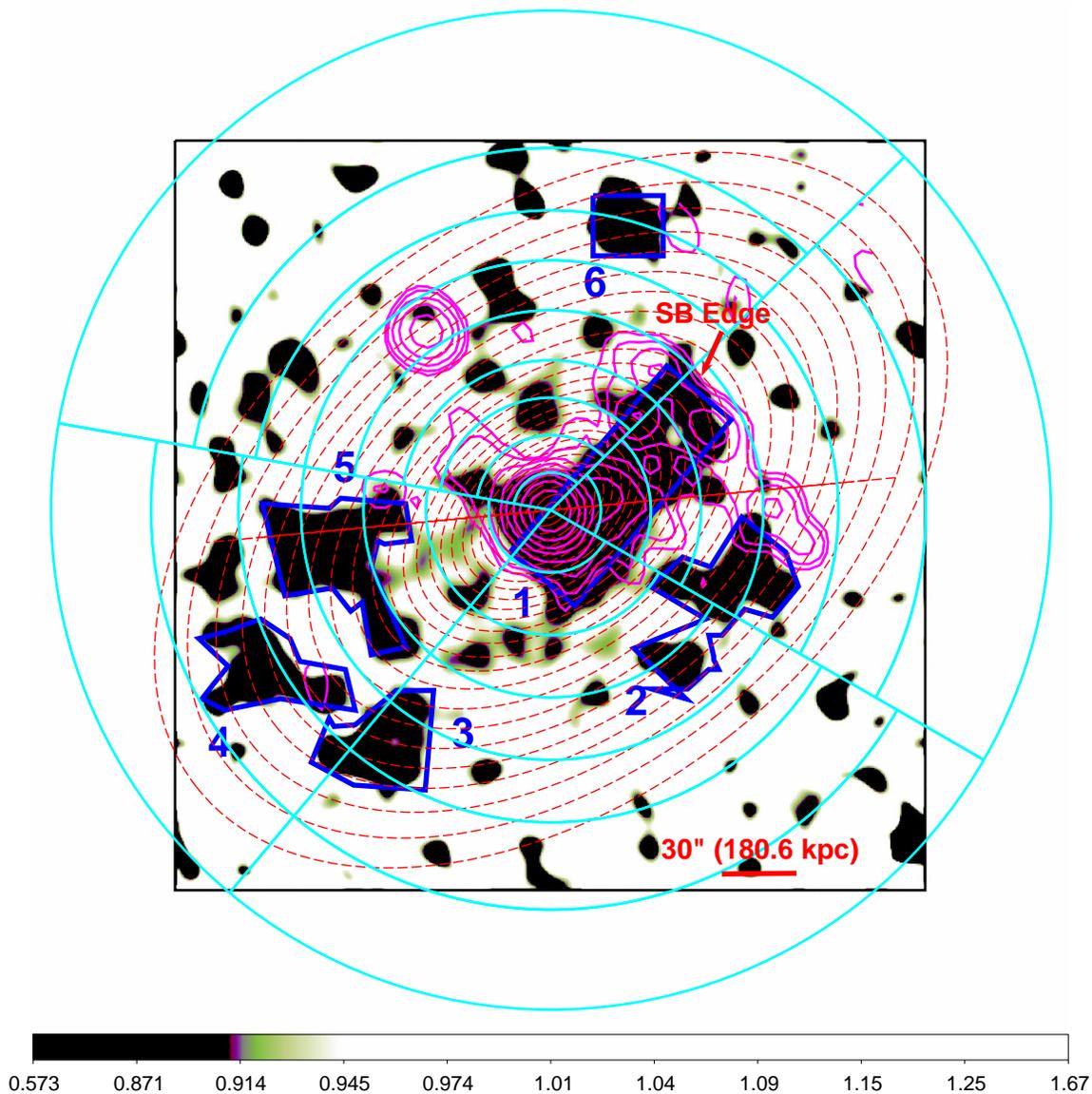}
}
\caption{X-ray hardness map created by dividing a 1.5-8.0\,keV image by that of 0.3-1.5\,keV. Before division, 
both the images were smoothed with a Gaussian kernel of 10 pixel (5 arcsec). Dark shades in this figure represent the 
regions of cool ICM. The significant cool regions marked by blue sectors were excluded in deriving the temperature 
profile (Figure~\ref{fig:KT1_Z1} {\it top left}). Overlaid magenta contours represent the 1.4 GHz VLA L-band radio 
emission from the radio source associated with the cluster. Notice asymmetry in the radio contours, and its association 
with the cool ICM. The edge in the radio emission in the north-west direction is found to coincides with the edge seen 
in the X-ray image. Cyan colour regions indicates regions used in sector wise temperature profile.}
\label{fig:KT1_cool}
\end{figure*}

\begin{figure*}
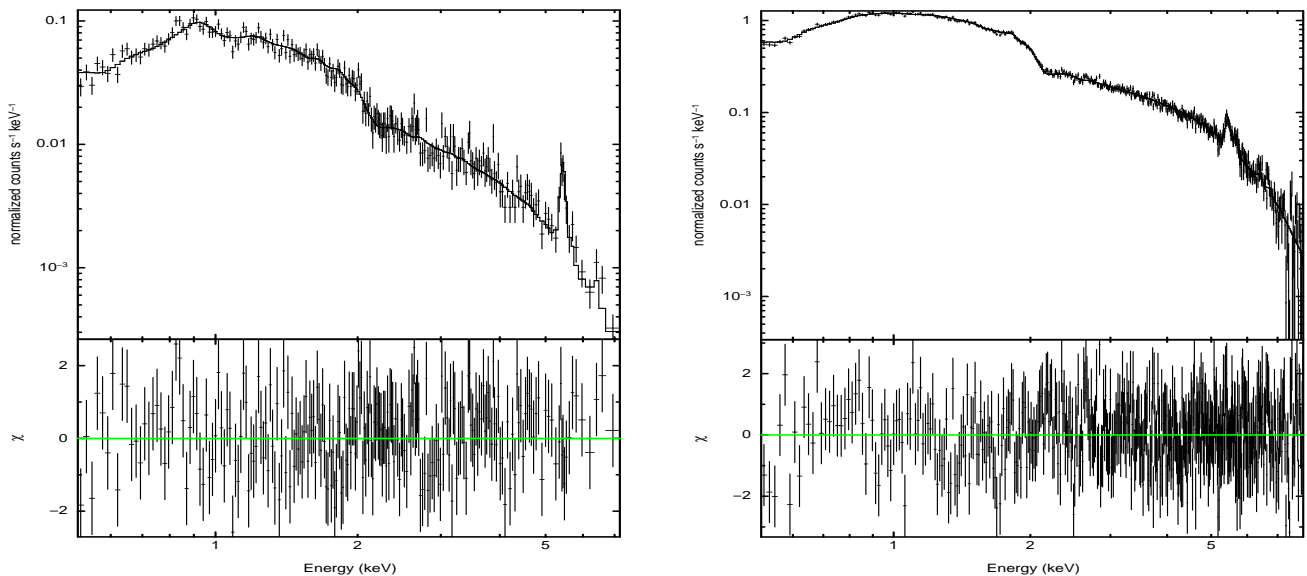

\vbox
{
\hspace*{-1.0cm}
\includegraphics[width=90mm,height=80mm]{fig9a.eps}
\includegraphics[width=90mm,height=80mm]{fig9b.eps}
}
\caption{{(\it left panel):} 0.5-7\,keV cumulative spectrum extracted from the central clumps. Notice the 6.4\,keV Fe line feature. {(\it right panel):} \textit{Chandra} 0.5 -7\,keV diffuse spectrum  extracted from within the 120 arcsec elliptical region along with the best fit absorbed two temperature APEC model.}
\label{fig:globla120} 
\end{figure*}

\subsection{Temperature and metallicity maps}

To trace the local variations of temperature and metal abundance of the ICM, we create two-dimensional temperature and abundance maps using the 95\,ks \textit{Chandra} data. To achieve this, X-ray photons were extracted from maps created using the \texttt{CONTBIN} adaptive binning algorithm described by \citet{Sanders06}. This algorithm locates the brightest pixels in the point source removed X-ray image and generates spatial bin around it including all neighbouring pixels with same brightness till the user-defined S/N threshold is not meet, which then moves outward and repeat the same exercise. The minimum S/N threshold was set to 60 ($\sim$ 3600 counts). Shape of each of the bin was restricted such that its length is at most two times of its width. This resulted in a spatially binned image which follows the surface brightness distribution \citep{Tremblay12}. 0.5-7\,keV source spectra, corresponding background, and associated response files were extracted for each of the region. The source spectra were then grouped to have a minimum of 15 counts in each bin. Each spectrum was then fitted with an absorbed thermal plasma model {\sc apec}, letting the temperature, metallicity and normalization to vary. The resulting temperature and metallicity maps for Abell 2390 are shown in Figure~\ref{fig:cont_maps} and reveals structures in both. The colourbars in this figure give the best fit temperature and abundance values in units of keV and $Z_\odot$, respectively.
The average temperature of the ICM on the north-western direction is statistically lower than that on the south-eastern direction. The large-scale temperature structure of Abell 2390 reveals its association with a cool core at the centre, where temperature drops to just 4\,keV relative to its virial value of 10\,keV.

Similar structures are also apparent in the metallicity map. The central abundance in Abell 2390 is roughly 0.54\,$Z_\odot$ which then decreases with radius until it reaches $\sim$0.15$Z_{\odot}$. Regions of enhanced metallicity are seen along the north and south direction at about 20\arcsec\,(72\,kpc) on either directions. This region is associated with the temperature jump at 20\arcsec\, evident in Figure~\ref{fig:KT1_Z1} {(\it top left \& top right)}. A comparison of the temperature and metallicity maps of Abell 2390 reveals that the regions of the high ICM temperature corresponds to poor metallicites, and are in good agreement with those reported by \citet{Baldi07} and \citet{Vikhlinin05}. Low metallicity wider regions one on the north-east and the other on the south-west direction are evident.

\subsection{Cluster Diffuse Emission}
To determine the weighted average properties of the diffuse X-ray emission we extracted and fitted a combined spectrum for the total diffuse emission from within an elliptical region with semi-major axis 120\arcsec\,($\sim$ 434\,kpc) region. We excluded the central 1.\arcsec5 region as well as the regions of obvious point sources. This yielded a total of 178,445 background-subtracted counts in the energy range between 0.5$-$7\,keV, which were then binned so as to have atleast 40 counts per bin. We initially fitted the spectrum with an absorbed single temperature {\sc apec} model, letting the temperature, abundance and normalization to vary and fixing the absorption at the Galactic value \citep{Vikhlinin05}. The resulting model yielded the best fit average temperature and abundance equal to $kT$ = 8.27$\pm$0.09\,keV and $Z$ = 0.33$\pm$0.02\,$\Zsun$, respectively, where the errors corresponds to 68\% confidence level. This single temperature APEC fit models the gas adequately at larger radii but gives poor fit in the inner region below 1\,keV. This is indicative of the relatively low temperature gas at the centre of the cluster. To account for this cooler component, a second thermal component was added to the  original model with column density fixed. The two-temperature model provided a significant improvement in the fit with the best fit temperatures equal to 9.56$^{+0.25}_{-0.19}$\,keV and 0.82$^{+0.18}_{-0.08}$\,keV at $\chi^2$/dof = 498.86/498. The best fit spectrum is shown in Figure~\ref{fig:globla120} {(\it right panel)}, while the resultant parameters are listed in Table~\ref{spectralpro}. We quantified the unabsorbed X-ray luminosity using the LUMIN function equal to $L_{X}$  =  $2.68\pm 0.01\times$10$^{45}$ erg\,s$^{-1}$. 

\subsection{Nuclear Point Source}
\textit{Wavdetect} algorithm detected a central X-ray point source in the cluster that coincides with  the optical core of the cD galaxy as well as the radio source in the 1.4\,GHz radio image of VLA. To investigate spectral properties of the nuclear source associated with Abell 2390, we extracted a 0.5-7\,keV spectrum from within the central  2\arcsec circular region, background for which was determined locally from an annulus surrounding the source. The spectrum was initially fitted with an absorbed power-law, which yielded into an acceptable fit with photon index $\Gamma\sim {\rm 1.86\pm0.06}$ and $\chi^2$/dof = 52.06/50.

We then tried with an additional absorption component intrinsic to the source, which resulted in a slightly better fit with $\chi^2$/dof = 49.72/49 and photon index $\Gamma\sim {\rm 2.0\pm0.11}$. This means the central source in Abell 2390, like that in other AGN cases, hosts sufficiently hard component. The additional absorbing column density was found to be equal to 6.87$\times$ 10$^{20}$ cm$^{-2}$ and is not significantly large relative to the Galactic value of 1.07$\times$ 10$^{21}$ cm$^{-2}$ \citep{Vikhlinin05}. From this thermal plus power-law model we estimate the unabsorbed 0.3-12\,keV flux of the central source equal to ${\rm 7.42\pm0.54\times 10^{-14}\,erg\,cm^{-2}\,s^{-1}}$, and the X-ray luminosity $L_X = {\rm 1.18\pm 0.07\times 10^{43}\,erg\,s^{-1}}$.}

\subsection{Central gas clumps}
In order to examine the spectral nature of the hot gas clump-like substructures apparent in the central region of Abell 2390 (Figure~\ref{fig:clumps}), we extract 0.5-7\,keV spectra from each of the clumpy substructure and fit with an absorbed single-temperature {\sc apec} model. The best fit parameters of the spectra from these three different clumpy regions reveal a systematic rise in temperature of the gas in clumps 1, 2, 3 (Table~\ref{spectralpro}) with highest temperature 5.21$^{+0.44}_{-0.37}$\,keV for clump 3. Existence of the gas clumps of systematically increasing temperatures and metallicity in the central region of this cluster suggests that Abell 2390 has passed through a merger episode and has not yet fully relaxed. Similar cases have also been reported in the past \citep[e.g. Abell 262;][]{blan04}. A combined 0.5-7\,keV spectrum of the X-ray photons from all the three clumpy regions was fitted with an absorbed {\sc apec} model is shown in Figure~\ref{fig:globla120} ({\it left panel}) and the best fit parameters are listed in Table~\ref{spectralpro}. 

\section{Discussion}
\label{disc}
\subsection{Cavity Energetics}
\begin{table*}
\caption{X-ray cavity parameters: size, age and energetics}
\begin{tabular}{@{}llllllcccccr@{}}
\hline\hline
{ Cavity Parameters}		&{E Cavity}		&{W Cavity} 	
&{N Cavity}		&{ S Cavity}	& {NW Cavity}	\\
\hline	
 $R_{1}$ (kpc)		&25.55			&20.01		
&37.54			&34.46			&12.88		\\
 $R_{w}$ (kpc)		&18.93  		&16.94		
&17.81			&16.79			&09.60		\\
 $D$ (kpc)		&85.07			&39.82		
&50.68			&51.40			&94.12		\\
 $t_{buoyancy}$(yr)	&2.87$\times10^8$	&1.04$\times10^8$
&1.09$\times10^8$	&2.89$\times10^8$	&4.73$\times10^8$	\\
 $P_{cav}$(erg\,s$^{-1}$)&7.10$\times10^{44}$	&1.58$\times10^{45}$
&2.72$\times10^{45}$	&8.6$\times10^{44}$	&7.2$\times10^{43}$	\\
\hline
\end{tabular}
\footnotesize
\begin{flushleft} 
{Note} : $R_{1}$, $R_{w}$ - semi-major axes along and perpendicular to the jet direction, respectively, 
D represents the projected distance from the core to cavity centre; $t_{buoyancy}$ - age of the cavities estimated using buoyancy rise; and $P_{cav}$ - cavity power using buoyancy age.
\end{flushleft}
\label{cavitypro}
\end{table*}

X-ray cavities or depressions in the surface brightness distribution are devoid of gas at the local ambient temperature and are believed to be formed due to the interactions between the AGN outbursts and the hot ICM. During an AGN outburst, the jets emanating from the central source do {\it pV} work on the surrounding ICM, inflate cavities which then rises buoyantly in the plane of the sky. Thus, it is possible to quantify the amount of mechanical energy that has been injected by the jets into the ICM from the analysis of the X-ray cavities \citep{Bir04,Rafferty06}. In other words, the X-ray cavities act as a calorimeter  to estimate the mechanical power injected by the AGN in to the ICM without requiring radio observations of the central source. 

Systematic analysis of the 95\,ks \textit{Chandra} data on Abell 2390 enabled us to detect a total of five X-ray cavities in the central 30\arcsec\,region. Assuming that each cavity takes spheroidal or oblate sphere geometry with its minor axis in the plane of the sky and filled with the material of negligible mass \citep{Randall11}, we estimate power of the cavities ($P_{cav}$) as the ratio of the energy contained by the cavities to their ages. The amount of energy required to inflate a cavity with pressure $p$ and volume $V$ is its enthalpy and is defined as sum of the internal energy $E$ of the cavity and work done ($pV$ ) by the jet to displace the X-ray emitting gas while it inflates the radio lobes i.e.,

\begin{equation}
\hspace{10mm} E_{cav} = H = E + pV = \frac{\gamma}{\gamma -1} pV ,
\end{equation}

where $\gamma$ is the mean adiabatic index of the fluid within the bubble and for the case of non-relativistic plasma it is 5/3. For non-relativistic content the total cavity energy is $E_{cav} = 2.5pV$ \citep{Mc07}.

The gas pressure $p$ of the surrounding ICM and volume V of the cavities were estimated directly from the analysis of the high resolution X-ray data \citep[see][]{Rafferty06}. Accuracy of the power estimation rely on the uncertainties in the measurement of volume of the X-ray cavities. As volume measurement is generally done by the visual inspection of the cavities in the X-ray image, therefore, is prone to systematic errors and is highly dependent on the quality of the X-ray data. The projection effect is also important in determining the spatial geometry of the cavities. As a result, volumes of the cavities measured by different observers may vary significantly due to their different approaches \citep{Gitti10} and hence may lead to the inaccurate estimation of the cavity power. To overcome these uncertainties, we estimate the cavity size from the surface brightness profile of the central region of the cluster. The extents of the cavities were measured from the decrements in the surface brightness profiles extracted from different sectors (Figure~\ref{fig:azmuth_beta} {\it right panel}). The resulting profiles show dip at the location of the cavity. X-ray cavities are assumed to be prolate ellipsoids. Volume of each of the cavity was estimated as $V = 4 \pi R_{1} R_{w}^2 /3$, where $R_{1}$ and $R_{w}$ are the projected semi-major axes along
and perpendicular to the jet direction, respectively.  Locations and distances of cavities were measured using the unsharp mask image. Our estimates of cavity volumes are 3.68$\times$10$^{69}$ cm$^{3}$, 2.35$\times$ 10$^{69}$ cm$^{3}$, 4.85$\times$10$^{69}$ cm$^{3}$, 3.96$\times$ 10$^{69}$ cm$^{3}$, and 4.83$\times$10$^{68}$ cm$^{3}$, respectively, for the E, W, N, S and NW cavities.} The temperature, density and pressure of the ICM surrounding the cavities ($p=1.92\,n_e k T$) were taken from the projected radial profiles of the temperature and electron density computed above at the radial distances corresponding to the centre of the cavities. Table~\ref{cavitypro} summarizes the cavity properties and their derived energetics.

Then the rate at which energy is injected into the ICM by the central AGN  was computed by dividing the cavity energy by the bubble age as
\begin{equation}
\hspace{25mm} P_{cavity}  \approx  \frac{E_{total}} { t_{age}} ,
\end{equation}

Here age of the cavity was estimated using the buoyancy rise method \citep{Chur01}. The buoyancy rise time is the time taken by the cavity to reach its terminal velocity, which depends on the drag forces. In the case of clearly detached cavities, the buoyancy rise method provides the better estimate of the cavity age and is given by 

\[ t_{buoyancy} \approx  R \sqrt{\frac{A\,C_D}{2gV}} \] 

Here, R is the projected distance from the centre of the cavity to the X-ray centroid of the ICM, $A$ = $\pi \times R_{w}^{2}$ is the cross-sectional area of the cavity, $C_D = 0.75$ is the drag coefficient \citep{Chur01}, and $g=2\sigma^2/R$ is the local gravitational acceleration. Our estimates of the age of the cavities using the buoyancy rise method listed in Table~\ref{cavitypro} were used to quantify their power content. The net power supplied by the central AGN in the form of the mechanical energy is found to be equal to 5.94$\times 10^{45}$ erg\,s$^{-1}$ and is lower than 1.0$^{+1.0}_{-0.9}\times 10^{46}$ erg\,s$^{-1}$ reported by \citet{Russell13} and \citet{hla11}. The total mechanical power content of the Abell 2390 is higher than that estimated in several other systems (e.g. NGC 6338, Abell 1991, IRAS 09104+4109, RBS 797, HCG 62, ZwCl 2701, Abell 262 etc.) and therefore is an important candidate for its detailed investigation. 
This estimate may even enhance by several factors if we include the power due to the cold fronts and other features apparent in this system \citep[e.g.][]{Fab03, Fab06, Form05}. Thus, the cavity power estimation using the X-ray analysis provides a lower limit to the true mechanical power of the AGN, i.e. the true jet power $P_{jet} \geq P_{cav} = E_{cav}/t_{age}$. Our estimate of the total mechanical power for all the five cavities $P_{cav} = L_{mech} = {\rm 5.94\times 10^{45}\,erg\,s^{-1}}$ and is roughly higher by a factor of $\sim$ 14 than the cooling luminosity $ L_{cool} {\rm= 4.04\pm0.03 \times 10^{44}\,erg\,s^{-1}}$ within the cooling radius.

\begin{figure*}
{
\includegraphics[width=80mm,height=80mm]{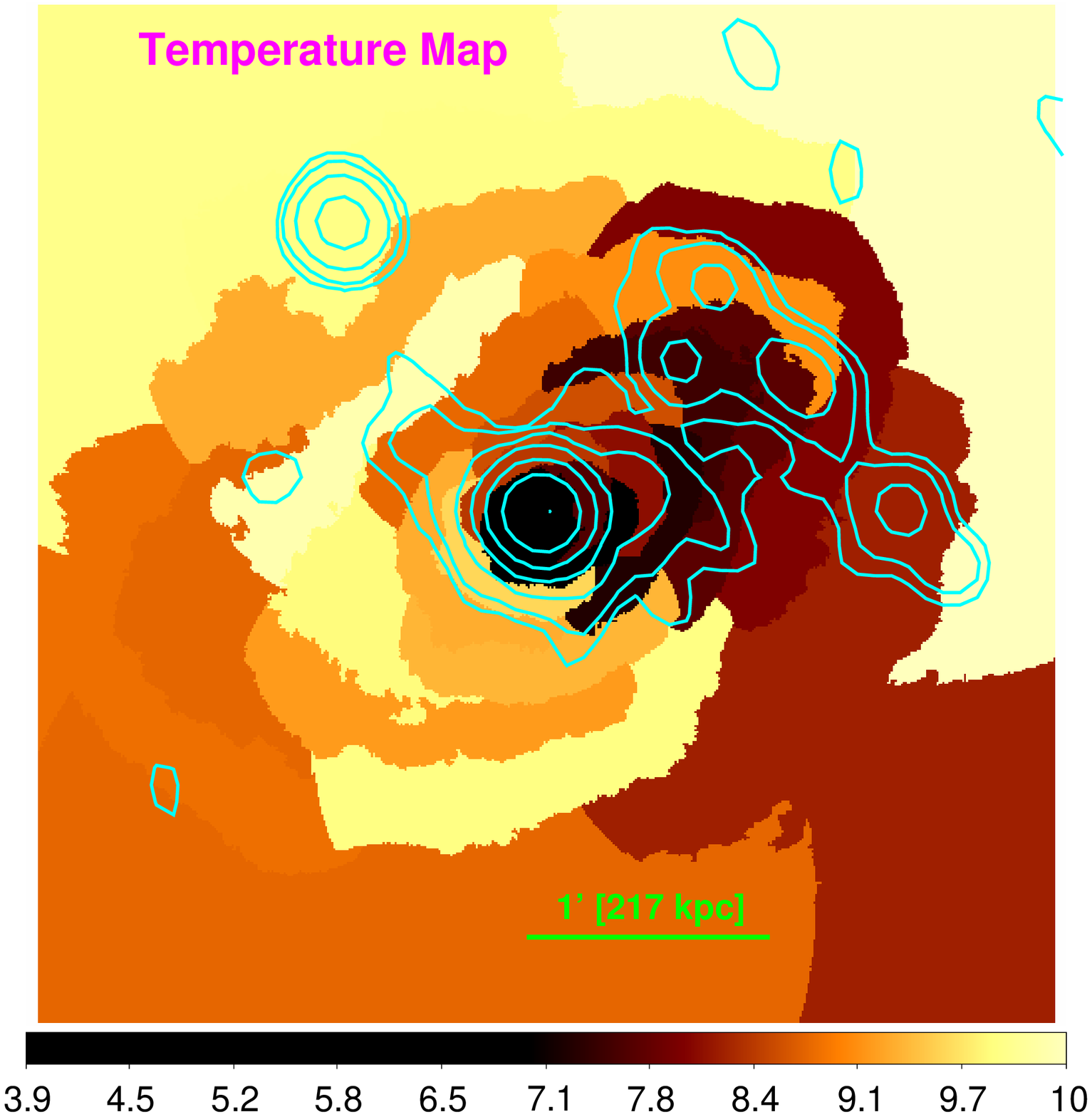}
\hspace{0.5cm}
\includegraphics[width=80mm,height=80mm]{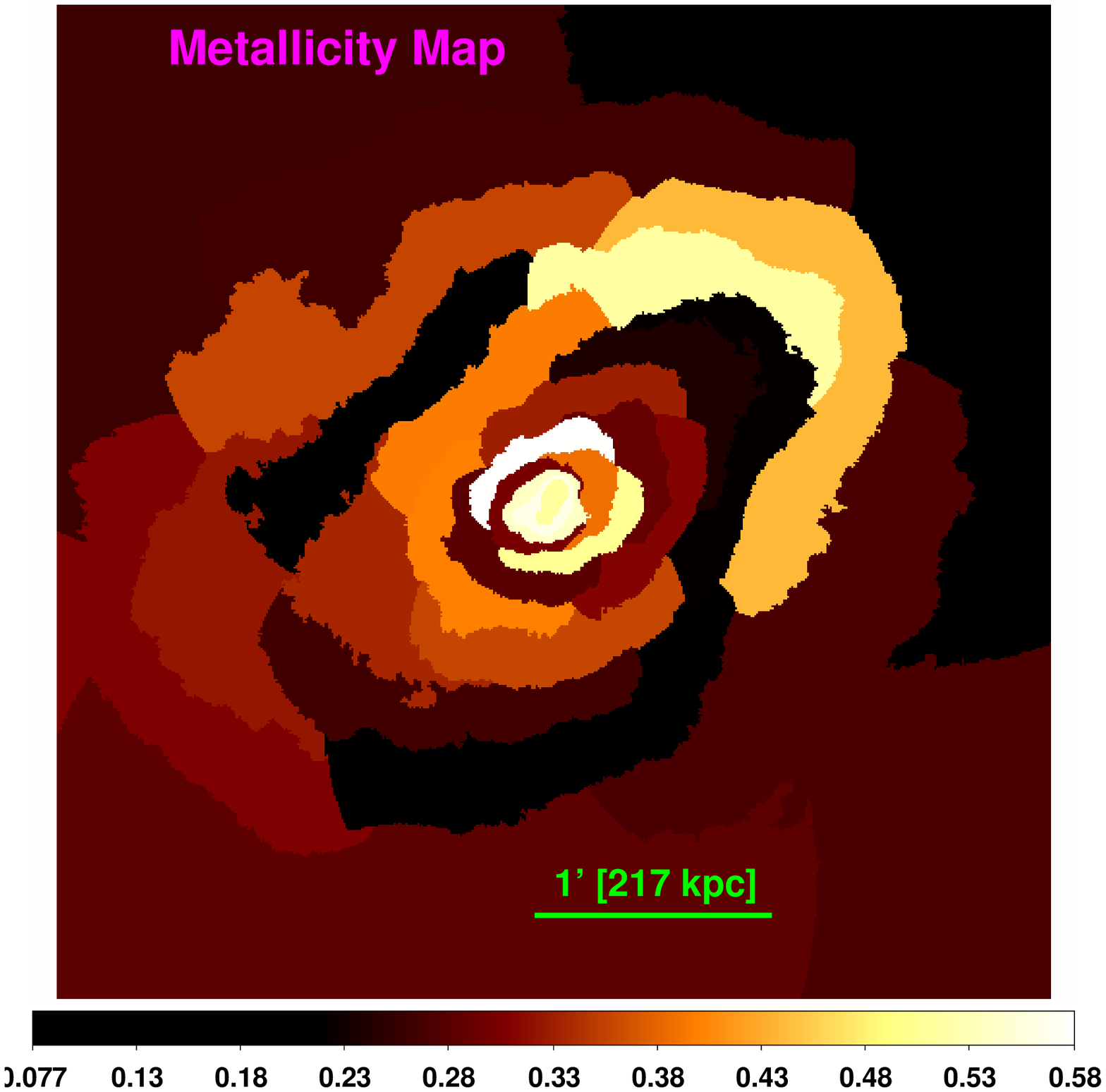}
}
\caption{Temperature ({\it left panel}) and metallicity ({\it right panel}) maps computed by grouping regions of similar surface brightness using the contour binning method. Temperature of the ICM on the south-east direction is statistically higher than that in the north-west direction. The temperature jumps at central 20\arcsec\, apparent in circular regions could be due to the cold fronts. Metallicity map reveals structures, such that metallicity on the north-west direction is relatively high compared to that on the south-east direction.}
\label{fig:cont_maps}
\end{figure*}

\subsection{Comparison with the radio data}
\begin{figure*}
\hbox{
\hspace{-1.0cm}
\includegraphics[width=120mm,height=120mm]{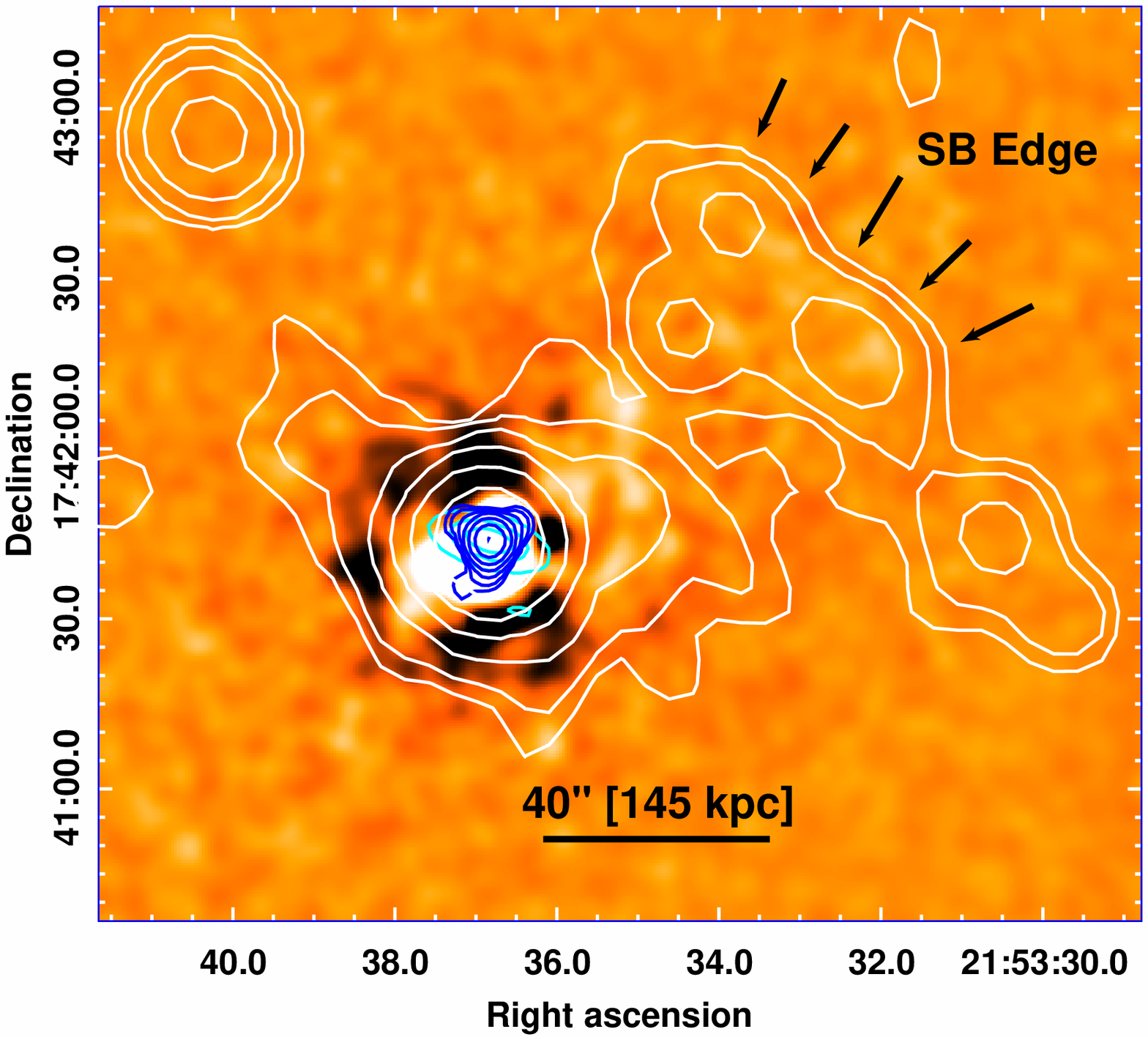}
}
\caption{1.4\,GHz VLA L-band radio contours (white) overlaid on 0.5$-$3\,keV \textit{Chandra} unsharp-masked image of Abell 2390. The diffuse radio emission map at 1.4 GHz exhibits highly irregular and asymmetric morphology of the central radio source with an elongation in the cavity direction. The extension along the north-west and north-east direction is associated with the relatively cool gas in the temperature map. The complex edge seen in radio emission along the north-west direction is found to be associated with the X-ray edge in the \textit{Chandra} unsharp mask image. We also plot the radio contours at 4.8\,GHz (blue) and 8.4\,GHz  (cyan) using the VLA data in the same figure, and are to be confined to the central 2\arcsec\,region only. The beam size at 1.4\,GHz is 14.36 $\times$ 14.36 arcsec and the rms noise is 126 $\mu$Jy $beam^{-1}$. The logarithmically spaced L-band radio contours are at 0.44, 0.87, 2.24, 6.58, 20.30, 63.67 and 200.84 mJy $beam^{-1}$.}
\label{fig:unsharp_radio1}
\end{figure*}

X-ray cavities in the ICM are thought to be inflated by radio jets originating from the AGN, that rise buoyantly until they reach density equilibrium with the surrounding ICM. Therefore, with an objective to examine association of the core of the Abell 2390 with a radio source, we make use of the 1.4 GHz VLA L-band radio data available in the \textit{NRAO Science Data Archive}. The diffuse radio emission contours at 1.4 GHz from Abell 2390 from their analysis are overlaid on the 0.5-3\,keV \textit{Chandra} unsharp mask image (Figure~\ref{fig:unsharp_radio1}). The radio emission morphology exhibits a complex nature that nearly match with the X-ray. The strongest radio source at the centre coincides with the cluster dominant cD galaxy PGC 140982, while the  diffuse radio emission at 1.4 GHz is found to cover all the detected X-ray cavities. Though, its morphology is highly asymmetric. The radio emission also exhibits a sharp edge on the south-east direction, while another more complex edge on the north-west direction due to several discrete radio sources is evident in this figure \citep{Bac03}. This complex edge in radio emission is found to coincide with the X-ray edge evident in the \textit{Chandra} unsharp mask image (Figure~\ref{fig:unsharp_radio1}). The extended 1.4 GHz diffuse radio emission along the north-west and the south-east directions exhibit its spatial association with the cool ICM evident in the temperature map (Figure~\ref{fig:cont_maps} {\it left panel}). Association of the strong radio source with the cooling flow cluster Abell 2390 suggest that this cluster falls in the class of the radio mini-halo and is similar in size to that of the Perseus cluster \citep{Burns92}. Non-spherical and more irregular morphology of diffuse radio emission from Abell 2390 is similar to that seen in the mini-halo of A2626 \citep{Gitti04}. 
We also plot the radio emission contours at 4.8\,GHz (blue) and 8.4\,GHz (cyan) from the VLA data in the same figure. Emission at higher frequencies appears to be confined only to the central $\sim$ 20\arcsec\,region. 

\subsection{ICM heating by the AGN feedback}

\begin{figure*}
\vbox
{
\hspace*{-0.5cm}
\includegraphics[width=88mm,height=88mm]{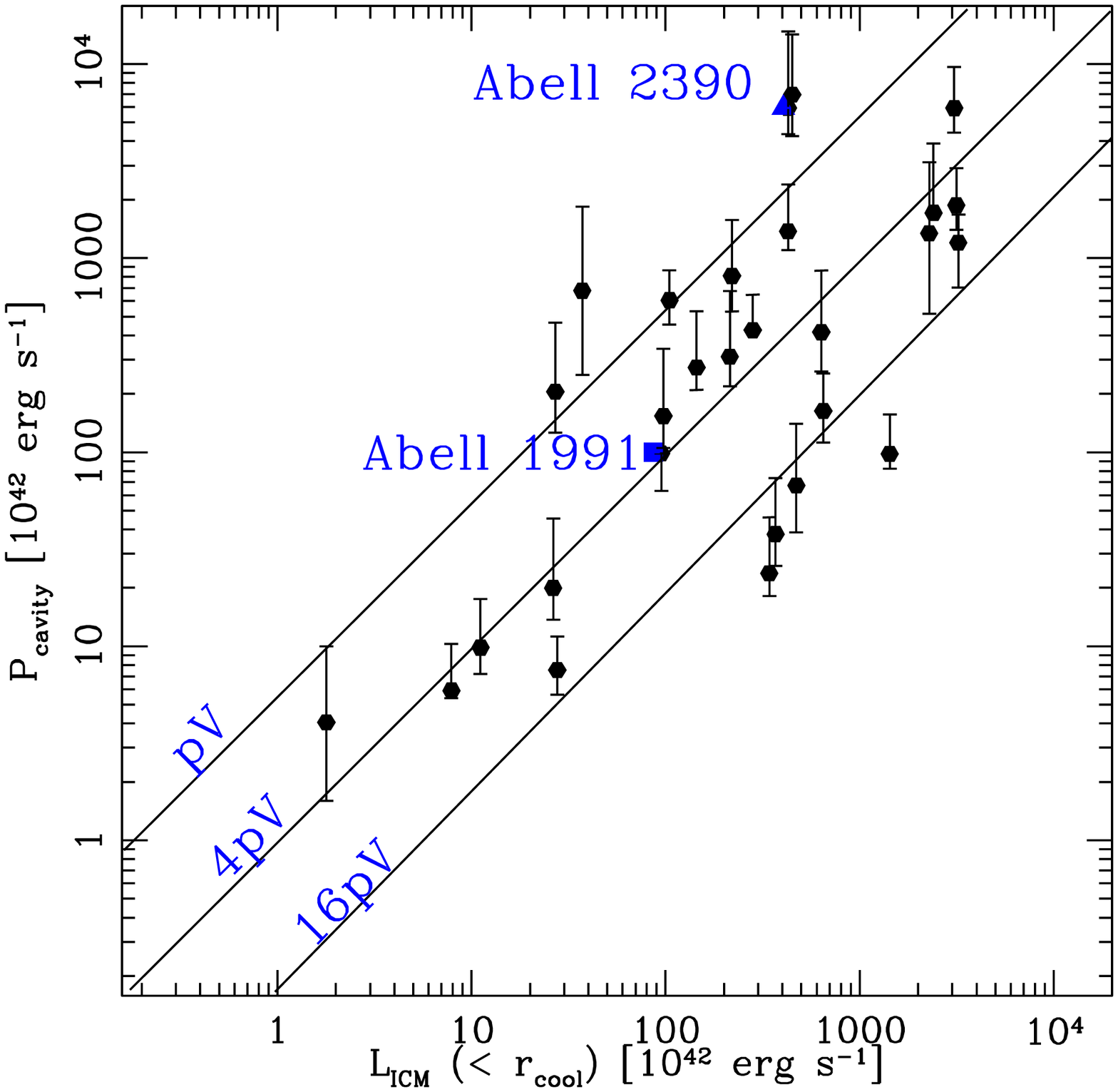}
\includegraphics[width=88mm,height=88mm]{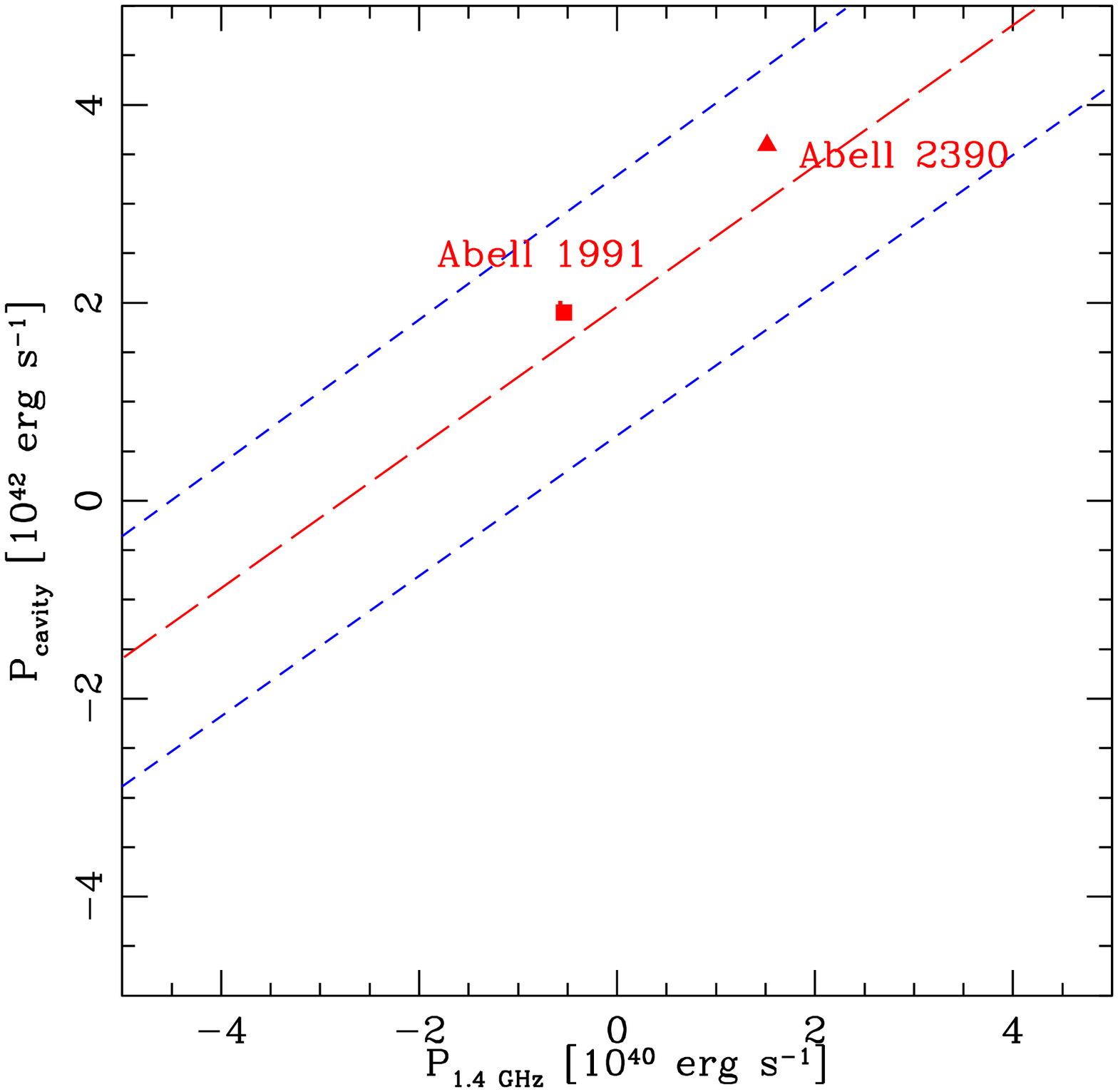}
}
\caption{({\it left}): Corelation between the cavity power $P_{cav}$ versus the X-ray luminosity of the ICM from within the cooling region ($L_{cool}$). The cavity power is estimated assuming the $4pV$ of the energy per cavity and their buoyancy rise time. The diagonal lines indicate their equivalence ($P_{cav} = L_{cool}$) at 1$pV$, 4$pV$ and 16$pV$, respectively, from top to bottom. Filled black symbols represent data points for the cluster sample of \citet{Cav10}, while the filled square denote Abell 1991 from \citep{Pan13}. The filled blue triangle represent the balance between the two for the case of Abell 2390. Error bars are at 1$\sigma$  uncertainties in cavity power estimation.   
({\it right}): Comparison between the cavity power ($P_{cav}$) versus 1.4 GHz radio power ($P_{1.4\,GHz}$) for the sample studied by \citet{Cav10}. The red long dashed line represent the best fit relation for their sample of giant ellipticals (gEs)  with the upper and lower limits shown by the short dashed blue lines. Abell 2390 occupies a position above the best-fit line by \citet{Cav10}.
}
\label{fig:pcav_lx} 
\end{figure*}

Radiative cooling of the ICM can be estimated by measuring X-ray luminosity of the hot gas within cooling radius of a cluster. In the present case for Abell 2390 it is found to be equal to $L_{cool} = {\rm 4.04\pm0.03 \times 10^{44}\,erg\,s^{-1}}$ corresponding to the radius of 60.91 kpc and cooling time of 1.9\,Gyr \citep{hla11}.
This means, in the absence of heating source, the cluster ICM would have radiated away all of its thermal energy over the time less than age of the cluster. However, estimate of the X-ray luminosity using deep X-ray data required some heating, e.g. AGN feedback,  to make up for the power radiated from within this region in order to suppress cooling.  This means, to balance the radiative cooling, the average energy input by the AGN outburst must be of the order or higher than the rate of radiative cooling. To examine efficiency of the AGN feedback, estimated mechanical power injected by the AGN from X-ray cavity analysis. $P_{cav}$ was compared with that of the radiative loss ($L_{cool}$). 

For the case of Abell 2390, we estimate the total X-ray cavity power of the central radio source to be $P_{cav} \sim$ 5.94$\times$ 10$^{45}$ erg s$^{-1}$ and has cavity enthalpy inexcess with that of the radiated power within cooling radius $L_{cool}$($ < r_{cool}$) $\sim$ 4.04$\times$ 10$^{44}$ erg s$^{-1}$ by almost an order of magnitude. To further elucidate this balance between the two,  we make use of the Figure 6 by \citet{Rafferty06}, where they plot the AGN power against the total radiative luminosity of the ICM within cooling radius (Figure~\ref{fig:pcav_lx} {\it left panel}). The diagonal lines in this plot indicate their equivalence ($P_{cavity} = L_{cool}$) at 1$pV$, 4$pV$ and 16$pV$ of total enthalpy (from top to bottom, respectively). Filled black symbols indicate data points for the cluster sample from \citet{Cav10}, while that indicated by filled square is for A1991 \citep{Pan13}. Error bars indicate the 1$\sigma$ uncertainties in the cavity power estimation. We also plot position of Abell 2390 in the same figure so as to check the balance between the two (filled triangle) and is found to occupy the position above the 1$pV$ line. This means, mechanical power fed by the central AGN associated with Abell 2390 is sufficiently large to inflate the cavities and to balance cooling of the ICM without 
requiring any additional contribution from the energy available with the relativistic particles and the magnetic field of the radio jets. 

\citet{Bir12} demonstrate that every massive cooling flow cluster with X-ray cavities hosts a radio source at its core with a radio luminosity 
greater than 2.5$\times$ 10$^{30}$ erg s$^{-1}$ Hz$^{-1}$. Following this, we explored the availability of radio data on this system and 
found that Abell 2390 is associated with a radio source of flux 235.3 mJy at 1.4\,GHz \citep{Cond98,Voi04}. 
The radio power of the central source associated with Abell 2390 was calculated using the relation \citep{Cav10}
\begin{equation}
P_{\nu_{0}}=4 \pi D_{\rm{L}}^{2} S_{\nu_{0}} \nu_{0} (1+z)^{\alpha-1},
\end{equation}
where $D_{\rm{L}}$, $S_{\nu_{0}}$, $z$ and $\alpha$ represent the luminosity distance, flux density at frequency $\nu_0$, redshift and the radio spectral index (assuming $S_{\nu} \sim \nu^{-\alpha}$), respectively. This lead to the radio power at 1.4\,GHz equal 
to $\sim$ 5.01$\times$ 10$^{41}$ erg s$^{-1}$. Thus, Abell 2390 meets the requirement of \citet{Bir12} and therefore increases 
the possibility of cavity heating of the ICM. We used this to confirm 
the balance between the two by plotting the mechanical power of cavities from X-ray study against the 1.4\,GHz radio power of the 
central radio source as suggested by \citet{Cav10} and is shown in Figure~\ref{fig:pcav_lx} ({\it right panel}). The long dashed red 
line in this figure represents the best fit relation obtained by \citet{Cav10} for a sample of giant ellipticals (gEs) with upper 
and lower limits shown by the short dashed blue lines. In this plot Abell 2390 is found to occupy position much above the best-fit line 
obtained by \citet{Cav10}, implying that our estimate of mechanical power using the X-ray data requires a much powerful radio source to meet the balance. Our estimate of the cavity power ($P_{cav} \sim$ 5.94$\times$ 10$^{45}$ erg s$^{-1}$) matches within error with 
that ($P_{cav} \sim$ 1.0$\pm$1.0 $\times$ 10$^{46}$ erg s$^{-1}$) reported by \cite{Russell13}. The measured values of radio 
flux densities at 1.4 GHz leads to the radio power equal to $\sim$ 5.01$\times$ 10$^{41}$ erg s$^{-1}$ with a ratio of the two 
powers ($P_{1.4GHz}$ / $P_{cav}$) equal to $\sim 10^{-4}$, which is lower compared to the theoretical range of $10^{-1}$ - $10^{-2}$
obtained by \cite{Bick97} as well as the reported values by \cite{Cav10} and \cite{Sull11}. This lower value is probably due to the uncertainties in the estimation of volume of the X-ray cavities.
As volume measurement is carried out by the visual inspection of the cavities in the X-ray image, therefore, is prone to systematic errors. According to \cite{Bir08} volume estimation as well as radio emission is also less sensitive to projection effect. 
If we consider total radio power over the frequency range 10 MHz to 10 GHz 
($\sim$ 3.67$\times$ 10$^{42}$ erg s$^{-1}$), this ratio turns out to be $\sim 10^{-3}$.
Similar values of the ratio of the two power estimates have also been reported for few other cases e.g. NGC 4636 \citep{Baldi09}, 
HCG 62 \citep{Gitti10} and few other sources \citep{Bir08}, where the mechanical power estimated from cavity analysis is 
found to exceed by about 4 orders of magnitude compared with the radio power. \cite{Gitti10} for the case of 
HCG 62, where $P_{1.4GHz}$ / $P_{cav}$ $\sim 10^{-4}$, have demonstrated that the synchrotron radio luminosity can not be taken 
as a reliable gauge of the total mechanical power of the AGN outburst. This is because the radio sources may be very poor, 
time-variable radiators or may fade with time. Another possibility is that the radio source associated with Abell 2390 is 
sufficiently powerful, whose power compares with that of Cygnus A, however, the denser environment of the cluster does not allow 
us to envisage the symmetrically placed lobes on either sides of the nucleus.

Another convincing probe for the AGN heating was provided by the nature of the entropy profile in central region of Abell 2390. 
The radial entropy profile derived for Abell 2390 falls systematically in the radially inward direction with a floor 
at 12.20$\pm$2.54\,keV cm$^2$ in the core region. If the cooling of the ICM is not compensated by the AGN heating, it would 
have radiated away whole of its thermal energy on the Hubble timescale and hence entropy would have collapsed to a very low 
value. However, the flattening of the entropy index towards the central projected value of 12.20$\pm$2.54\,keV cm$^2$ strongly 
point towards an intermittent heating of the ICM by the strong AGN feedback \citep{soker05}.

\citet{Raw12} using the \textit{Spitzer} and \textit{Herschel} data estimate the star formation rate in the core of this system 
to be equal to  $\sim$ 9 $\Msun$/yr. However, this estimate is significantly smaller than that expected by the standard cooling 
flow model, implying that the AGN is regulating it through the feedback process. They also found a strong correlation 
between the infrared luminosity and the cluster X-ray gas cooling time, suggesting that the star formation in this system is strongly 
influenced by the cluster-scale cooling process. Thus, all the attempts made here to relate the jet/cavity power with the X-ray/radio 
luminosity imply that the AGN outburst from the core of the Abell 2390 is capable enough to quench the cooling and also to 
carve the morphology of the ICM.

\section{Conclusions}
\label{conc}
In this paper we have presented systematic analysis of 95\,ks  high resolution \textit{Chandra} data on a cooling 
flow cluster Abell 2390 with an objective to investigate properties of the X-ray cavities apparent in the surface 
brightness distribution. 1.4 GHz VLA L-band data reveals an association of a strong radio source with the cD galaxy 
PGC 140982 at the core of this cluster. Comparison of the X-ray and radio data exhibit a strong interplay between 
the central radio source and the ICM, indicating that the cool core cluster Abel 2390 is an important source to 
investigate the AGN feedback. Main results from this study are summarized below:

\begin{itemize}
\item Our morphological analysis using a total of 95\,ks \textit{Chandra} data confirms the presence of a pair of X-ray cavities predicated by \citet{Allen01} and \citet{Vikhlinin05}. Additionally, this study has detected three more X-ray cavities in the central 30\arcsec\,region. The presence of these cavities have been confirmed by a variety of image processing techniques i.e., unsharp mask image, 2D elliptical-$\beta$ model subtracted residual image as well as by the surface brightness profiles derived along four different sectors.
\item Projected radial temperature profile reveals a positive temperature gradient in the central region, like those seen in several other cooling flow clusters. 
\item Temperature profiles derived along the four different wedge-shaped sectors revealed temperature jump in the all directions at a projected distance of $\sim$25\arcsec\,and another jump from 7.47\,keV to 9.10\,keV in the west direction at a projected distance of 68\arcsec\,(246 kpc). These jumps in temperature are due to the cold fronts and their corresponding Mach numbers are 1.44$\pm$0.05 and 1.22$\pm$0.06.
\item Tricolour map as well as hardness ratio map detects cool gas clumps in the central 30 kpc region  of temperature $4.45_{-0.10}^{+0.16}$\,keV. 
The cooler gas has a remarkable impact on the radial temperature profile of the cluster. After removing the contribution of the cool gas, Abell 2390's temperature profile is consistent with the form observed in other cool core galaxy clusters.
\item The entropy profile derived from the X-ray analysis is found to fall systematically inward in a power-law fashion and exhibits a floor near 12.20$\pm$2.54\,keV\, cm$^2$ in the central region. This flattening of the entropy profile in the core region confirms the intermittent heating at the centre by AGN. 
\item The diffuse radio emission map at 1.4\,GHz using VLA L-band data exhibits highly asymmetric morphology with an edge in the north-west direction coinciding with the X-ray edge seen in the unsharp mask image.
\item The mechanical power injected by the AGN in the form of X-ray cavities is found to be 5.94$\times$10$^{45}$ erg\,s$^{-1}$ and is roughly an order of magnitude higher than the energy lost by the ICM in the form of X-ray emission, confirming that AGN feedback is capable enough to quench the cooling flow in this cluster.

\end{itemize}

\section{Acknowledgements}
We thank the anonymous referee for useful comments and suggestions, that helped us to improve the quality of the manuscript greatly. 
SSS acknowledges financial support from the Ministry of Minority Affairs, Govt. of India, under the Minority Fellowship 
Program (award No. F-17.1/2010/MANF-BUD-MAH-2111/SA-III/Website). PKP acknowledges financial support from the CSIR, India. 
The authors gratefully acknowledge the use of computing and library facilities of the Inter-University Centre for Astronomy 
and Astrophysics (IUCAA), Pune, India and the High Performance Computing (HPC) facility procured under 
the DST, New Delhi's FIST scheme (F No. SR/FST/PS-145/2009). This work has made use of data from the 
\textit{Chandra} archive, NASA’s Astrophysics Database System (ADS), NASA/IPAC Extragalactic Database (NED), 
High Energy Astrophysics Science Archive Research Center (HEASARC). We thank Jeremy Sanders for his help on deriving the contour binned
maps and usage of the Veusz package.

\def\aj{AJ}%
\def\actaa{Acta Astron.}%
\def\araa{ARA\&A}%
\def\apj{ApJ}%
\def\apjl{ApJ}%
\def\apjs{ApJS}%
\def\ao{Appl.~Opt.}%
\def\apss{Ap\&SS}%
\def\aap{A\&A}%
\def\aapr{A\&A~Rev.}%
\def\aaps{A\&AS}%
\def\azh{AZh}%
\def\baas{BAAS}%
\def\bac{Bull. astr. Inst. Czechosl.}%
\def\caa{Chinese Astron. Astrophys.}%
\def\cjaa{Chinese J. Astron. Astrophys.}%
\def\icarus{Icarus}%
\def\jcap{J. Cosmology Astropart. Phys.}%
\def\jrasc{JRASC}%
\def\mnras{MNRAS}%
\def\memras{MmRAS}%
\def\na{New A}%
\def\nar{New A Rev.}%
\def\pasa{PASA}%
\def\pra{Phys.~Rev.~A}%
\def\prb{Phys.~Rev.~B}%
\def\prc{Phys.~Rev.~C}%
\def\prd{Phys.~Rev.~D}%
\def\pre{Phys.~Rev.~E}%
\def\prl{Phys.~Rev.~Lett.}%
\def\pasp{PASP}%
\def\pasj{PASJ}%
\def\qjras{QJRAS}%
\def\rmxaa{Rev. Mexicana Astron. Astrofis.}%
\def\skytel{S\&T}%
\def\solphys{Sol.~Phys.}%
\def\sovast{Soviet~Ast.}%
\def\ssr{Space~Sci.~Rev.}%
\def\zap{ZAp}%
\def\nat{Nature}%
\def\iaucirc{IAU~Circ.}%
\def\aplett{Astrophys.~Lett.}%
\def\apspr{Astrophys.~Space~Phys.~Res.}%
\def\bain{Bull.~Astron.~Inst.~Netherlands}%
\def\fcp{Fund.~Cosmic~Phys.}%
\def\gca{Geochim.~Cosmochim.~Acta}%
\def\grl{Geophys.~Res.~Lett.}%
\def\jcp{J.~Chem.~Phys.}%
\def\jgr{J.~Geophys.~Res.}%
\def\jqsrt{J.~Quant.~Spec.~Radiat.~Transf.}%
\def\memsai{Mem.~Soc.~Astron.~Italiana}%
\def\nphysa{Nucl.~Phys.~A}%
\def\physrep{Phys.~Rep.}%
\def\physscr{Phys.~Scr}%
\def\planss{Planet.~Space~Sci.}%
\def\procspie{Proc.~SPIE}%
\let\astap=\aap
\let\apjlett=\apjl
\let\apjsupp=\apjs
\let\applopt=\ao
\nocite{*}

\bibliographystyle{spr-mp-nameyear-cnd}


\begin{thebibliography}{0}
\ifx \bisbn   \undefined \def \bisbn  #1{ISBN #1}\fi
\ifx \binits  \undefined \def \binits#1{#1} \fi
\ifx \bauthor  \undefined \def \bauthor#1{#1} \fi
\ifx \batitle  \undefined \def \batitle#1{#1} \fi
\ifx \bjtitle  \undefined \def \bjtitle#1{#1}\fi
\ifx \bvolume  \undefined \def \bvolume#1{\textbf{#1}}\fi
\ifx \byear  \undefined \def \byear#1{#1} \fi
\ifx \bissue  \undefined \def \bissue#1{#1} \fi
\ifx \bfpage  \undefined \def \bfpage#1{#1} \fi
\ifx \blpage  \undefined \def \blpage #1{#1} \fi
\ifx \burl  \undefined \def \burl#1{\textsf{#1}} \fi
\ifx \doiurl  \undefined \def \doiurl#1{\textsf{#1}} \fi
\ifx \betal  \undefined \def \betal{\textit{et al.}} \fi
\ifx \binstitute  \undefined \def \binstitute#1{#1} \fi
\ifx \binstitutionaled  \undefined \def \binstitutionaled#1{#1} \fi
\ifx \bctitle  \undefined \def \bctitle#1{#1} \fi
\ifx \beditor  \undefined \def \beditor#1{#1} \fi
\ifx \bpublisher  \undefined \def \bpublisher#1{#1} \fi
\ifx \bbtitle  \undefined \def \bbtitle#1{#1} \fi
\ifx \bedition  \undefined \def \bedition#1{#1} \fi
\ifx \bseriesno  \undefined \def \bseriesno#1{#1} \fi
\ifx \blocation  \undefined \def \blocation#1{#1} \fi
\ifx \bsertitle  \undefined \def \bsertitle#1{#1} \fi
\ifx \bsnm \undefined \def \bsnm#1{#1} \fi
\ifx \bsuffix \undefined \def \bsuffix#1{#1} \fi
\ifx \bparticle \undefined \def \bparticle#1{#1} \fi
\ifx \barticle \undefined \def \barticle#1{#1} \fi
\ifx \bconfdate \undefined \def \bconfdate #1{#1} \fi
\ifx \botherref \undefined \def \botherref #1{#1} \fi
\ifx \url \undefined \def \url#1{\textsf{#1}} \fi
\ifx \bchapter \undefined \def \bchapter#1{#1} \fi
\ifx \bbook \undefined \def \bbook#1{#1} \fi
\ifx \bcomment \undefined \def \bcomment#1{#1} \fi
\ifx \oauthor \undefined \def \oauthor#1{#1} \fi
\ifx \citeauthoryear \undefined \def \citeauthoryear#1{#1} \fi
\ifx \endbibitem  \undefined \def \endbibitem {}\fi
\ifx \bconflocation  \undefined \def \bconflocation#1{#1} \fi
\ifx \arxivurl  \undefined \def \arxivurl#1{\textsf{#1}} \fi

\end{thebibliography}


\begin{thebibliography}

\bibitem[Abraham et al. (1996)]{Abr96} Abraham, R. G., Smecker-Hane, T. A., Hutchings, J. B., et al., 1996, \apj, 471, 694
\bibitem[Allen et al. (1998)] {Allen98} Allen, S. W. \& Fabian, A. C., 1998, \mnras, 297, L63
\bibitem[Allen et al. (2001)] {Allen01} Allen, S. W., Ettori, S., \& Fabian, A. C., 2001, \mnras, 324, 877
\bibitem[Arnaud et al. (1996)] {Arnaud96} Arnaud, K. A., 1996, in Astronomical Society of the Pacific Conference Series, Vol. 101, Astronomical Data Analysis Software and Systems V, Jacoby, G. H. \& Barnes, J., eds., p. 17
\bibitem[Augusto et al. (2006)] {Augusto06} Augusto, P., Edge, A. C., \& Chandler, C. J., 2006, \mnras, 367, 366
\bibitem[Bacchi et al. (2003)]{Bac03} Bacchi, M., Feretti, L., Giovannini, G., \& Govoni, F. 2003, A\&A, 400, 465
\bibitem[Baldi et al. (2007)] {Baldi07} Baldi A., Ettori S., Mazzotta P., Tozzi P., Borgani S., 2007, \apj, 666, 835
\bibitem[Baldi et al. (2009)] {Baldi09}	Baldi A., Forman W., Jones C., Kraft R., Nulsen P., Churazov E., David L., Giacintucci S., 2009, \apj, 707, 1034
\bibitem[Bardeau et al. (2007)]{Bardeau07} Bardeau, S., Soucail, G., Kneib, J.-P., Czoske, O., Ebeling, H., Hudelot, P., Smail, I., \& Smith, G. P., 2007, A\&A, 470, 449
\bibitem[Bicknell et al. (1997)]{Bick97} Bicknell, G. V., Dopita, M. A., \& O'Dea, C. P. 1997, \apj, 485, 112
\bibitem[{B{\^i}rzan} et al. (2004)]{Bir04} {B{\^i}rzan}, L., Rafferty, D.A., McNamara, B.R., Wise, M.W., Nulsen, P.E.J. 2004 , \apj, 607, 800 
\bibitem[{B{\^i}rzan} et al. (2008)]{Bir08} {B{\^i}rzan}, L., McNamara, B. R., Nulsen, P. E. J., Carilli, C. L., \& Wise, M. W., 2008, \apj, 686, 859
\bibitem[{B{\^i}rzan} et al. (2012)]{Bir12} {B{\^i}rzan}, L., Rafferty, D.A., Nulsen, P.E.J., McNamara, B.R., R¨ottgering, H.J.A., Wise,M.W., Mittal, R. 2012, \mnras, 427, 3468
\bibitem[Blanton et al. (2001)]{blan01} Blanton, E. L., Sarazin, C. L., McNamara, B. R., \& Wise, M. W., 2001, \apj, 558, L15
\bibitem[Blanton et al. (2003)]{blan03} Blanton, E., Sarazin, C. \& McNamara 2003, \apj, 585, 227
\bibitem[Blanton et al. (2004)]{blan04} Blanton, E. L., Sarazin, C. L., McNamara, B. R., \& Clarke, T. E., 2004, \apj, 612, 817
\bibitem[Blanton et al. (2009)]{blan09} Blanton, E. L., Randall, S. W., Douglass, E. M., Sarazin, C. L., Clarke, T. E.,\& McNamara, B. R. 2009, \apj, 697, L95
\bibitem[Blanton et al. (2011)]{blan11} Blanton E. L., Randall S. W., Clarke T. E., Sarazin C. L., McNamara B. R., Douglass E. M., McDonald M., 2011, \apj, 737, 99
\bibitem[Burns et al. (1992)] {Burns92}Burns, J. O., Sulkanen, M. E., Gisler, G. R., \& Perley, R. A. 1992, \apj, 388, L49
\bibitem[Canning et al. (2013)]{Can13} Canning, R. E. A., Sun, M., Sanders, J. S., et al., 2013, \mnras, 435, 1108
\bibitem[Cavagnolo et al. (2009)]{Cav09} Cavagnolo, K. W., Donahue, M., Voit, G. M., \& Sun, M. 2009, \apjs, 182, 12
\bibitem[Cavagnolo et al. (2010)]{Cav10} Cavagnolo, K.W., McNamara, B.R., Nulsen, P.E.J., Carilli, C.L., Jones, C., {B{\^i}rzan}, L. 2010, \apj,720,1066 
\bibitem[Churazov et al. (2001)]{Chur01}Churazov E., {Br{\"u}ggen} M., Kaiser C. R., {B{\"o}hringer} H., Forman W., 2001, \apj, 554, 261
\bibitem[Condon et al. (1998)]{Cond98} Condon, J. J., Cotton, W. D., Greisen, E. W., Yin, Q. F., Perley, R. A., Taylor, G. B., \& Broderick, J. J., 1998, \aj, 115, 1693
\bibitem[David et al. (2009)]{David09} David, L. P., Jones, C., Forman, W., Nulsen, P., Vrtilek, J., O'Sullivan, E., Giacintucci, S., \& Raychaudhury, S., 2009, \apj, 705, 624
\bibitem[Dong et al. (2010)]{dong10} Dong, R., Rasmussen, J., \& Mulchaey, J. S., 2010, \apj, 712, 883
\bibitem[Egami et al. (2006)]{Egami06} Egami, E., Misselt, K. A., Rieke, G. H., et al., 2006, \apj, 647, 922 
\bibitem[Fabian (2012)] {Fab12} Fabian, A. C., 2012, \araa, 50, 455
\bibitem[Fabian et al. (2003)]{Fab03} Fabian, A. C., Sanders, J. S., Allen, S. W., Crawford, C. S., Iwasawa, K., Johnstone, R. M., Schmidt, R. W., \& Taylor, G. B., 2003, \mnras, 344, L43
\bibitem[Fabian et al. (2006)]{Fab06} Fabian, A.C., Sanders, J.S., Taylor, G.B., Allen, S.W., Crawford, C.S., Johnstone, R.M.,Iwasawa, K. 2006, \mnras, 366, 417 
\bibitem[Forman et al. (2005)]{Form05} Forman, W., Nulsen, P., Heinz, S., et al., 2005, \apj, 635, 894
\bibitem[Gastaldello et al. (2007)]{Gastaldello07} Gastaldello, F., Buote, D. A., Humphrey, P. J., Zappacosta, L., Bullock, J. S., Brighenti, F., \& Mathews,W. G., 2007, \apj, 669, 158
\bibitem[Giacintucci et al. (2011)]{Giac11} Giacintucci, S., O'Sullivan, E., Vrtilek, J., et al., 2011, \apj, 732, 95
\bibitem[Gitti et al. (2004)]{Gitti04} Gitti, M., \& Schindler, S. 2004, A\&A, 427, L9
 \bibitem[Gitti et al. (2010)]{Gitti10} Gitti, M., O'Sullivan, E., Giacintucci, S., David, L. P., Vrtilek, J., Raychaudhury, S., \& Nulsen, P. E. J., 2010, \apj, 714, 758
\bibitem[Gitti et al. (2011)]{Gitti11} Gitti, M., Nulsen, P. E. J., David, L. P., McNamara, B. R., \& Wise, M. W. 2011, \apj, 732, 13
\bibitem[Gitti et al. (2012)]{Gitt12} Gitti, M., Brighenti, F., \& McNamara, B.R. 2012, Adv. Astron., 2012, 1
\bibitem[Haines et al. (2012)]{Haines12} Haines, C. P., Pereira, M. J., Sanderson, A. J. R., et al., 2012, \apj, 754, 97
\bibitem[Hicks et al. (2006)]{Hicks06} Hicks, A. K., Ellingson, E., Hoekstra, H., \& Yee, H. K. C., 2006, \apj, 652, 232
\bibitem[Hlavacek-Larrondo et al. (2011)]{hla11} Hlavacek-Larrondo, J. \& Fabian, A. C., 2011, \mnras, 413, 313
\bibitem[Mazzotta et al. (2003)]{Mazz03} Mazzotta, P., Edge, A. C., \& Markevitch, M., 2003, \apj, 596, 190
\bibitem[McNamara \& Nulsen (2007)]{Mc07} McNamara, B.R., Nulsen, P.E.J. 2007, \araa, 45, 117
\bibitem[McNamara et al. (2005)]{McN05} McNamara, B. R., Nulsen, P. E. J., Wise, M. W., Rafferty, D. A., Carilli, C., Sarazin, C. L., \& Blanton, E. L., 2005, \nat, 433, 45
\bibitem[Nulsen et al. (2005)]{Nul05}Nulsen P. E. J., McNamara B. R., Wise M. W., David L. P., 2005, \apj, 628, 629
\bibitem[O'Sullivan et al. (2011)]{Sull11} O'Sullivan E., Giacintucci S., David L. P., Gitti M., Vrtilek J. M., Raychaudhury S., Ponman T. J., 2011, \apj, 735, 11
\bibitem[Pandge et al. (2012)]{Pan12} Pandge, M.B., Vagshette, N.D., David, L.P., Patil, M.K. 2012, \mnras, 421, 808 
\bibitem[Pandge et al. (2013)]{Pan13} Pandge, M. B., Vagshette, N. D., Sonkamble, S. S., \& Patil, M. K., 2013, Ap\&SS, 345, 183
\bibitem[Pipino et al. (2009)]{Pipino09} Pipino, A., Kaviraj, S., Bildfell, C., Babul, A., Hoekstra, H., \& Silk, J., 2009, \mnras, 395, 462
\bibitem[Rafferty et al. (2006)]{Rafferty06} Rafferty, D. A., McNamara, B. R., Nulsen, P. E. J., \& Wise, M. W., 2006, \apj, 652, 216
\bibitem[Randall et al. (2011)]{Randall11} Randall, S.W., Forman, W.R., Giacintucci, S., Nulsen, P.E.J., Sun, M., Jones, C., Churazov, E.,David, L.P., Kraft, R., Donahue, M., Blanton, E.L., Simionescu, A., Werner, N. 2011, \apj, 726,86  
\bibitem[Rasmussen et al. (2007)]{Rasmussen07} Rasmussen, J. \& Ponman, T. J., 2007, \mnras, 380, 1554
\bibitem[Rawle et al. (2012)]{Raw12} Rawle, T. D., Edge, A. C., Egami, E., et al., 2012, ApJ, 747, 29
\bibitem[Russell et al. (2013)] {Russell13} Russell, H. R., McNamara, B. R., Edge, A. C., Hogan, M. T., Main, R. A., \& Vantyghem, A. N., 2013, \mnras, 432, 530
\bibitem[Sanders et al. (2006)] {Sanders06} Sanders, J. S., 2006, \mnras, 371, 829
\bibitem[Sanders et al. (2001)] {Sanders01} Sanders, J. S. \& Fabian, A. C., 2001, \mnras, 325, 178
\bibitem[Sharma et al. (2004)] {Sha04}Sharma, M., McNamara, B.R., Nulsen, P.E.J., Owers, M., Wise, M.W., Blanton, E.L., Sarazin,C.L., Owen, F.N., David, L.P. 2004, \apj, 613, 180 
\bibitem[Soker et al. (2005)]{soker05} Soker N., Pizzolato F., 2005, \apj, 622, 847
\bibitem[Sun et al. (2009)]{Sun09} Sun, M., Voit, G. M., Donahue, M., Jones, C., Forman, W., \& Vikhlinin, A., 2009, \apj, 693, 1142
\bibitem[Tremblay et al. (2012)] {Tremblay12} Tremblay, G. R., O'Dea, C. P., Baum, S. A., et al., 2012, \mnras, 424, 1026
\bibitem[Vikhlinin et al. (2005)] {Vikhlinin05} Vikhlinin, A., Markevitch, M., Murray, S. S., Jones, C., Forman,W., \& Van Speybroeck, L., 2005, \apj, 628, 655
\bibitem[Vikhlinin et al. (2006)] {Vikhlinin06} Vikhlinin, A., Kravtsov, A., Forman, W., Jones, C., Markevitch, M., Murray, S. S., \& Van Speybroeck, L., 2006, \apj, 640, 691
\bibitem[Voigt \& Fabian (2004)]{Voi04} Voigt, L.M., Fabian, A.C. 2004, \mnras, 347, 1130
\bibitem[Wise et al. (2004)]{wise04} Wise, M. W., McNamara, B. R., \& Murray, S. S. 2004, \apj, 601, 184
\bibitem[Yee et al. (1996)]{Yee96} Yee, H. K. C., Ellingson, E., Abraham, R. G., Gravel, P., Carlberg, R. G., Smecker-Hane, T. A., Schade, D., \& Rigler, M., 1996, \apjs, 102, 289

\end{thebibliography}
\bibliographystyle{spr-mp-nameyear-cnd}

\end{document}